\RequirePackage[OT1]{fontenc}
\documentclass[10pt, journal]{IEEEtran}

\usepackage{amsmath,amsfonts}
\usepackage{graphicx}
\usepackage{cite}
\usepackage{makecell}
\usepackage{siunitx}
\usepackage[acronym]{glossaries}
\usepackage{xurl}
\usepackage{algorithm}
\usepackage[noend]{algpseudocode}
\usepackage{hyperref}
\usepackage{mathtools}
\usepackage{orcidlink}
\usepackage{booktabs}
\usepackage{microtype}
\usepackage{tikz}
\usetikzlibrary{fit,calc}

\ifCLASSOPTIONcompsoc
  \usepackage[caption=false,font=normalsize,labelfont=sf,textfont=sf]{subfig}
\else
  \usepackage[caption=false,font=footnotesize]{subfig}
\fi

\newcommand*{\tikzmk}[1]{\tikz[remember picture,overlay] \node (#1) {};\ignorespaces}
\newcommand{\boxit}[4]{\tikz[remember picture,overlay]{\node[yshift=#4,fill=#1,opacity=.25,fit={(A)($(B)+(#2,#3)$)},blend mode=darken] {};}\ignorespaces}

\DeclareSIUnit{\nothing}{\relax}

\makeatletter
\def\bstctlcite{\@ifnextchar[{\@bstctlcite}{\@bstctlcite[@auxout]}}
\def\@bstctlcite[#1]#2{\@bsphack
  \@for\@citeb:=#2\do{%
    \edef\@citeb{\expandafter\@firstofone\@citeb}%
    \if@filesw\immediate\write\csname #1\endcsname{\string\citation{\@citeb}}\fi}%
  \@esphack}
\makeatother

\newacronym{2d}{2D}{two-dimensional}

\newacronym{asr}{ASR}{automatic speech recognition}
\newacronym{nmt}{NMT}{neural machine translation}
\newacronym{tts}{TTS}{text-to-speech}
\newacronym{sid}{SID}{speaker identification}

\newacronym{brir}{BRIR}{binaural room impulse response}

\newacronym{dnn}{DNN}{deep neural network}
\newacronym{lstm}{LSTM}{long short-term memory}

\newacronym{ic}{IC}{interaural coherence}
\newacronym{ild}{ILD}{interaural level difference}
\newacronym{irm}{IRM}{ideal ratio mask}
\newacronym{itd}{ITD}{interaural time difference}

\newacronym{logfbe}{log-FBE}{log-filterbank energy}

\newacronym{mse}{MSE}{mean squared error}

\newacronym{pesq}{PESQ}{perceptual evaluation of speech quality}

\newacronym{relu}{ReLU}{rectified linear unit}
\newacronym{rms}{RMS}{root mean square}

\newacronym{snr}{SNR}{signal-to-noise ratio}
\newacronym{stoi}{STOI}{short-term objective intelligibility}
\newacronym{estoi}{ESTOI}{extended short-term objective intelligibility}
\newacronym{sisnr}{SI-SNR}{scale-invariant signal-to-noise ratio}

\newacronym{tf}{TF}{time-frequency}
\newacronym{stft}{STFT}{short-time Fourier transform}

\newacronym{zpr}{ZPR}{zero-padding rate}

\newacronym{sde}{SDE}{stochastic differential equation}
\newacronym{ve}{VE}{variance exploding}
\newacronym{vp}{VP}{variance preserving}

\newacronym{gan}{GAN}{generative adversarial network}
\newacronym{vae}{VAE}{variational autoencoder}
\newacronym{nf}{NF}{normalizing flow}

\newacronym{ema}{EMA}{exponential moving average}

\newacronym{pc}{PC}{predictor-corrector}

\newacronym{ou}{OU}{Ornstein-Uhlenbeck}

\newacronym{pdf}{PDF}{probability density function}

\glsenableentrycount

\def\dpesq{\Delta\text{\cgls{pesq}}}
\def\destoi{\Delta\text{\cgls{estoi}}}
\def\dsnr{\Delta\text{\cgls{snr}}}

\newcommand*\verythinspace{\mskip0.5\thinmuskip}
\newcommand*\verythinnegativespace{\mskip-0.5\thinmuskip}
\newcommand*\inlineeq{\!=\!}
\newcommand*\inlineplus{\!+\!}
\newcommand*\inlineminus{\!-\!}
\newcommand*\inlinein{\verythinnegativespace\in\verythinnegativespace}
\newcommand*\inlinelt{\!<\!}
\newcommand*\inlinesim{\!\sim\!}
\newcommand*\inlineleq{\!\leq\!}
\newcommand*\inlinegeq{\!\geq\!}

\newcommand{\myabovecaptionskip}{}
\newcommand{\mybelowcaptionskip}{}
\newcommand{\myabovesinglefigureskip}{}
\newcommand{\myabovedoublefigureskip}{}

\makeatletter
\def\smallunderbrace#1{\mathop{\vtop{\m@th\ialign{##\crcr
   $\hfil\displaystyle{#1}\hfil$\crcr
   \noalign{\kern3\p@\nointerlineskip}%
   \tiny\upbracefill\crcr\noalign{\kern1\p@}}}}\limits}
\makeatother

\newcommand{\squared}[2]{#1^2 \mskip-0.25\thinmuskip (#2)}

\newcommand*\diff{\mathop{}\!\mathrm{d}}

\newcommand*\wiener{\boldsymbol{\omega}_t}
\newcommand*\smin{\sigma_{\mathrm{min}}}
\newcommand*\smax{\sigma_{\mathrm{max}}}
\newcommand*\bmin{\beta_{\mathrm{min}}}
\newcommand*\bmax{\beta_{\mathrm{max}}}
\newcommand*\bx{\boldsymbol{x}}
\newcommand*\bz{\boldsymbol{z}}
\newcommand*\by{\boldsymbol{y}}
\newcommand*\bs{\boldsymbol{s}}
\newcommand*\bF{\boldsymbol{F}}
\newcommand*\bD{\boldsymbol{D}}
\newcommand*\bn{\boldsymbol{n}}
\newcommand*\bff{\boldsymbol{f}}
\newcommand*\bmu{\boldsymbol{\mu}}
\newcommand*\btheta{\boldsymbol{\theta}}

\newcommand*\bSigma{\boldsymbol{\Sigma}}
\newcommand*\sigmahat{\bar{\sigma}}
\newcommand*\bxhat{\bar{\bx}}
\newcommand*\bxtildehat{\bar{\bxtilde}}
\newcommand*\phat{\bar{p}}
\newcommand*\ptilde{\tilde{p}}
\newcommand*\ptildehat{\bar{\ptilde}}
\newcommand\perturbkernel[1]{p_{0t}(#1_t|#1_0)}
\newcommand\perturbkernelhat[2]{\phat_{0t}(#1_t|#2_0)}
\newcommand\perturbkernelsgmse{p_{0t}(\bx_t|\bx_0,\by)}
\newcommand*\score{\nabla_{\bx_t}\!\log p_t(\bx_t)}
\newcommand*\scoreconditioned{\nabla_{\bx_t}\!\log p_t(\bx_t | \by)}
\newcommand*\scorehat{\nabla_{\bx_t}\!\log \phat ( \bxhat_t, \sigmahat(t) )}
\newcommand*\scoretildehat{\nabla_{\bx_t}\!\log \ptildehat ( \bxtildehat_t, \by, \sigmahat(t) )}
\newcommand*\scoremodel{\bs_{\btheta}}
\newcommand*\model{\bF_{\btheta}}
\newcommand*\denoiser{\bD_{\btheta}}
\newcommand*\idealdenoiser{\bD}
\newcommand*\scalednoise{\bn_t}
\newcommand*\unscalednoise{\bar{\bn}_t}
\newcommand*\cskip{c_{\mathrm{skip}}}
\newcommand*\cin{c_{\mathrm{in}}}
\newcommand*\cout{c_{\mathrm{out}}}
\newcommand*\cshift{c_{\mathrm{shift}}}
\newcommand*\cnoise{c_{\mathrm{noise}}}
\newcommand*\bxtilde{\tilde{\bx}}
\newcommand*\pdata{p_{\mathrm{data}}}
\newcommand*\sdata{\sigma_{\mathrm{data}}}
\newcommand*\ptildedata{\ptilde_{\mathrm{data}, \by}}
\newcommand*\R{\mathbb{R}}
\newcommand*\E{\mathbb{E}}
\newcommand*\C{\mathbb{C}}
\newcommand*\gaussian{\mathcal{N}}
\newcommand*\uniform[2]{\mathcal{U}(#1, #2)}

\newcommand*\tansq{\tan^{2}\mskip-1.5\thinmuskip}
\newcommand*\cossq{\cos^{2}\mskip-1.5\thinmuskip}
\newcommand*\cotsq{\cos^{2}\mskip-1.5\thinmuskip}

\title{Investigating the Design Space of Diffusion Models for Speech Enhancement}

\author{
Philippe~Gonzalez\,\orcidlink{0009-0006-4965-3514},
Zheng-Hua~Tan\,\orcidlink{0000-0001-6856-8928},
Jan~{\O}stergaard\,\orcidlink{0000-0002-3724-6114},
Jesper~Jensen\,\orcidlink{0000-0003-1478-622X},
Tommy~Sonne~Alstr{\o}m\,\orcidlink{0000-0003-0941-3146},
Tobias~May\,\orcidlink{0000-0002-5463-5509}
}

\IEEEaftertitletext{\vspace{-1\baselineskip}}

\begin{document}

\robustify\bfseries

\bstctlcite{IEEEexample:BSTcontrol}

\maketitle

\begin{abstract}
Diffusion models are a new class of generative models that have shown outstanding performance in image generation literature.
As a consequence, studies have attempted to apply diffusion models to other tasks, such as speech enhancement.
A popular approach in adapting diffusion models to speech enhancement consists in modelling a progressive transformation between the clean and noisy speech signals.
However, one popular diffusion model framework previously laid in image generation literature did not account for such a transformation towards the system input, which prevents from relating the existing diffusion-based speech enhancement systems with the aforementioned diffusion model framework.
To address this, we extend this framework to account for the progressive transformation between the clean and noisy speech signals.
This allows us to apply recent developments from image generation literature, and to systematically investigate design aspects of diffusion models that remain largely unexplored for speech enhancement, such as the neural network preconditioning, the training loss weighting, the \cgls{sde}, or the amount of stochasticity injected in the reverse process.
We show that the performance of previous diffusion-based speech enhancement systems cannot be attributed to the progressive transformation between the clean and noisy speech signals.
Moreover, we show that a proper choice of preconditioning, training loss weighting, \cgls{sde} and sampler allows to outperform a popular diffusion-based speech enhancement system while using fewer sampling steps, thus reducing the computational cost by a factor of four.
\end{abstract}

\begin{IEEEkeywords}
Diffusion models, speech enhancement.
\end{IEEEkeywords}

\begin{tikzpicture}[remember picture,overlay]
\node[anchor=south,yshift=3pt] at (current page.south) {
  \fbox{\parbox{\dimexpr\textwidth - 2\fboxsep}{
    \footnotesize \copyright 2024 IEEE. Personal use of this material is permitted. Permission from IEEE must be obtained for all other uses, in any current or future media, including reprinting/republishing this material for advertising or promotional purposes, creating new collective works, for resale or redistribution to servers or lists, or reuse of any copyrighted component of this work in other works.
  }}
};
\end{tikzpicture}
\vspace{-\baselineskip}

\section{Introduction}
\label{sec:intro}

\IEEEPARstart{T}{he} presence of noise and reverberation in an acoustic scene can substantially reduce speech intelligibility, for both normal-hearing and hearing-impaired listeners~\cite{nabelek1974monaural,nabelek1981effect}.
It can also reduce the performance of downstream tasks, such as \cgls{asr}~\cite{palomaki2004techniques,palomaki2004binaural} and \cgls{sid}~\cite{may2011noise,may2012binaural}.
Therefore, speech enhancement techniques that aim to restore speech intelligibility and quality by removing additive and convolutional distortions from noise and reverberation are crucial for a wide range of applications, such as telecommunications and hearing aid technology.

For the past years, the vast majority of studies on speech enhancement have favored deep neural network-based techniques, due to their superior performance over traditional statistical methods~\cite{xu2015regression,wang2018supervised,bentsen2018benefit}.
Deep neural network-based systems can be categorized into discriminative and generative models.
Discriminative models learn a mapping from the input noisy speech to the output clean speech.
They are trained in a supervised manner by presenting them with noisy and clean speech signals in pairs.
On the other hand, generative models learn a probability distribution over clean speech.
This probability distribution can be conditioned on an input noisy speech signal to perform speech enhancement.
While discriminative models account for the vast majority of neural network-based speech enhancement systems~\cite{wang2018supervised}, generative models represent an attractive approach, as learning the inherent properties of speech should enable robustness to arbitrary additive and convolutional distortions.
Moreover, discriminative models have shown to produce unpleasant speech distortions that can be detrimental to downstream \cgls{asr} tasks~\cite{narayanan2014investigation,wang2019bridging}.
Generative approaches include \cglspl{gan}~\cite{goodfellow2014generative}, \cglspl{vae}~\cite{kingma2014auto} and \cglspl{nf}~\cite{rezende2015variational}, which have all been applied to speech enhancement~\cite{michelsanti2017conditional,bando2018statistical,nugraha2020flow}.

Diffusion models~\cite{sohl2015deep,song2021score,ho2020denoising} have recently gained significant interest for the task of speech enhancement~\cite{lu2021study,lu2022conditional,welker2022speech,richter2023speech,lemercier2023storm,yen2023cold,chen2023metric,wang2023cross,sawata2023versatile,qiu2023srtnet,tai2023dose,shi2024diffusion,gonzalez2024diffusion,nortier2024unsupervised,ayilo2024diffusion}.
As a new class of generative models, they have met great success in image generation~\cite{ho2020denoising,dhariwal2021diffusion,rombach2022high}, audio generation~\cite{kong2021diffwave,popov2021grad,liu2023audioldm} and video generation~\cite{ho2022video,ho2022imagen}.
They consist in simulating a diffusion process by progressively adding Gaussian noise to the training data until it can be approximated as pure Gaussian noise.
A neural network is then trained to undo this process and new samples can be generated by starting from new random noise realizations.
In~\cite{lu2021study,lu2022conditional}, a discrete Markov chain formulation inspired by~\cite{ho2020denoising} was adopted, where the mean of the diffusion process was a linear interpolation between the clean and noisy speech signals.
Speech enhancement was then performed by sampling from random noise centered around the noisy speech signal and solving the reverse diffusion process.

In~\cite{welker2022speech,richter2023speech}, a continuous \cgls{sde} formulation inspired by~\cite{song2021score} was proposed, where a novel drift term allowed for a progressive transformation of the mean of the diffusion process from the clean speech signal towards the noisy speech signal.
Compared to the discrete Markov chain formulation in~\cite{lu2021study,lu2022conditional}, a continuous setting is more general and allows to use an arbitrary numerical solver to integrate the reverse process.
It was also argued to allow for the derivation of an objective function that does not explicitly involve the difference between the clean speech signal and the noisy speech signal as opposed to~\cite{lu2021study,lu2022conditional}, thus making the training task purely generative.
The introduced drift term was also argued to enable the system to reconstruct speech corrupted by environmental noises that are different from the stationary Gaussian noise used in the diffusion process~\cite{richter2023speech}.

Another common approach to diffusion-based speech enhancement consists in combining the diffusion model with a discriminative model that predicts a first estimate of the clean speech signal~\cite{lemercier2023storm,wang2023cross,sawata2023versatile,qiu2023srtnet,shi2024diffusion}.
The diffusion model is then tasked with removing unwanted distortions introduced by the discriminative model using e.g.\ denoising diffusion restoration models~\cite{kawar2022denoising,sawata2023versatile}, stochastic regeneration~\cite{lemercier2023storm} or stochastic refinement~\cite{whang2022deblurring,qiu2023srtnet}.
This approach avoids typical generative artifacts such as vocalizing and breathing effects, and considerably reduces the computational overhead from the diffusion model, since the number of steps in the reverse diffusion process can be reduced without losing performance~\cite{lemercier2023storm}.
Finally, another promising approach consists in using an unconditional diffusion model to solve inverse problems~\cite{kawar2022denoising,chung2022improving,chung2023diffusion}.
This was successfully applied to different audio applications such as music restoration~\cite{moliner2023solving}, speech dereverberation~\cite{lemercier2023diffusion} and source separation~\cite{iashchenko2023undiff}.
This approach paves the way for a new class of diffusion-based speech enhancement models that can be trained in an unsupervised way.

Despite these recent advances, several aspects of diffusion-based speech enhancement models remain poorly understood.
In particular, the role of the drift term that pulls the mean of the diffusion process towards the noisy speech in~\cite{welker2022speech,richter2023speech} is unclear.
In our view, the noisy speech signal can be seen as a conditioner for the speech enhancement task in the same way as a text prompt in image generation, the only difference being that in speech enhancement, the conditioner has the same modality as the training data.
Since such a drift is not needed to obtain good results in image generation, it is unclear why the success of diffusion-based speech enhancement should be attributed to it.
Moreover, it was shown by Karras et al.~\cite{karras2022elucidating} that multiple diffusion models that initially differed in their formulation can be framed in a common framework, allowing to better compare them.
However, due to the nature of the drift term used for speech enhancement in~\cite{welker2022speech,richter2023speech}, the resulting diffusion model cannot be directly included in the framework from~\cite{karras2022elucidating}.
As a consequence, it is difficult to translate recent advances from image generation literature to speech enhancement.
Finally, while some design aspects of diffusion-based speech enhancement were analyzed in subsequent works~\cite{lemercier2023analysing,lay2023reducing}, many remain unexplored, such as the neural network preconditioning~\cite{karras2022elucidating}, the loss weighting, the \cgls{sde}, or the amount of stochasticity injected when integrating the reverse process.

Our contributions in this work are as follows:
\begin{itemize}
    \item We extend a diffusion model framework previously laid in image generation literature~\cite{karras2022elucidating} to include a recently proposed diffusion-based speech enhancement system~\cite{welker2022speech,richter2023speech} where the long-term mean of the diffusion process is non-zero and equal to the conditioner.
    Compared to~\cite{gonzalez2024diffusion} where we used a change of variable to consider the \cgls{sde} satisfied by the environmental noise and fit in the framework laid in~\cite{karras2022elucidating}, we instead propose here a more general approach that models the clean speech signal and accounts for an eventual drift of the diffusion process.
    \item Using this new framework, we show that the success of diffusion-based speech enhancement models cannot be attributed to the addition of a drift term that pulls the mean of the diffusion process towards the conditioner.
    This also proves that a mismatch between the final distribution of the forward process and the prior distribution used to initialize the reverse process is not necessarily responsible for a drop in performance as suggested in~\cite{lay2023reducing}.
    \item We systematically investigate the design space of diffusion models for speech enhancement by experimenting with different neural network preconditionings~\cite{karras2022elucidating}, loss weightings, \cglspl{sde}, and amounts of stochasticity injected when integrating the reverse process.
    \item This investigation results in a system that outperforms a popular diffusion-based baseline while using fewer sampling steps, thus allowing for a fourfold decrease in computational cost.
\end{itemize}

In Sect.~\ref{sec:related}, we go over some related work on diffusion models and their adaptation to speech enhancement.
In Sect.~\ref{sec:extending}, we prove how the framework from~\cite{karras2022elucidating} can be extended to include the diffusion-based speech enhancement system from~\cite{welker2022speech,richter2023speech}.
In Sect.~\ref{sec:experimental}, we describe the experimental setup.
In Sect.~\ref{sec:results}, we present and discuss the results.
Code is available online\footnote{\url{https://github.com/philgzl/brever}}.

\section{Related work}
\label{sec:related}

\subsection{Diffusion models}
\label{sec:diffusion_models}

A diffusion model progressively transforms clean training data ${\bx_0 \inlinein \R^d}$ with distribution ${\pdata}$ into isotropic Gaussian noise.
This transformation can be modeled as a diffusion process ${\bx_t \inlinein \R^d}$ indexed by a continuous variable ${t \inlinein [0, \verythinnegativespace 1]}$ satisfying the following general \cgls{sde},
\begin{equation}
    \label{eq:general_sde}
    \diff \bx_t = \bff(t, \bx_t) \diff t + g(t) \diff \wiener,
\end{equation}
where ${\bff(t, \cdot) : \R^d \to \R^d}$ is the \textit{drift} coefficient, ${g : \R \to \R^+}$ is the \textit{diffusion} coefficient and ${\wiener \inlinein \R^d}$ is a standard Wiener process.
The goal of a diffusion model is to solve the associated reverse-time \cgls{sde}~\cite{anderson1982reverse}, such that new samples can be generated starting from random Gaussian noise realizations at ${t \inlineeq 1}$.
The reverse-time \cgls{sde} associated with Eq.~\eqref{eq:general_sde} is
\begin{equation}
    \label{eq:reverse_sde}
    \diff \bx_t = \left[ \bff(t, \bx_t) - \squared{g}{t} \score \right] \diff t + g(t) \diff \wiener,
\end{equation}
where ${\score}$ is termed the \textit{score function}, which is intractable as it involves marginalizing over the unknown distribution of the training data ${\pdata}$,
\begin{equation}
    p_t(\bx_t) = \int_{\R^d} \perturbkernel{\bx} \pdata(\bx_0) \diff \bx_0,
\end{equation}
where the conditional \cgls{pdf} ${p_{0t}}$ is commonly referred to as the \textit{perturbation kernel}.
Even though ${\score}$ is intractable, a neural network can be trained to approximate it via \textit{score matching}~\cite{hyvarinen2005estimation,vincent2011connection,song2020sliced},
\begin{equation}
    \scoremodel(\bx_t, t) \simeq \score,
\end{equation}
where ${\scoremodel}$ is the score model.
New samples can then be generated from random Gaussian noise realizations by substituting ${\score}$ with ${\scoremodel(\bx_t, t)}$ in Eq.~\eqref{eq:reverse_sde} and integrating Eq.~\eqref{eq:reverse_sde} using an arbitrary numerical solver (e.g.\ Euler-Maruyama or Runge-Kutta methods).
The number of sampling steps ${n_{\mathrm{steps}}}$ used to discretize the time axis and integrate the reverse process controls the number of neural network evaluations.
Therefore, ${n_{\mathrm{steps}}}$ defines a compromise between sample quality and computational cost.

The drift coefficient is commonly chosen to have the form ${\bff(t, \bx_t) \inlineeq f(t) \verythinspace \bx_t}$, such that the \cgls{sde} in Eq.~\eqref{eq:general_sde} becomes
\begin{equation}
    \label{eq:linear_sde}
    \diff \bx_t = f(t) \verythinspace \bx_t \diff t + g(t) \diff \wiener.
\end{equation}
The perturbation kernel of this \cgls{sde} turns out to be Gaussian and can be calculated using Eqs.~(5.50) and (5.51) in~\cite{sarkka2019applied},
\begin{equation}
    \label{eq:perturb_ker}
    \perturbkernel{\bx} = \gaussian \big( \bx_t ; s(t) \verythinspace \bx_0, \squared{\sigma}{t} \verythinspace \mathbf{I} \big),
\end{equation}
where
\begin{equation}
    \label{eq:marginals}
    \resizebox{.91\hsize}{!}{$
    \displaystyle
    s(t) = \exp \int_0^t f(\xi) \diff \xi
    \quad \text{and} \quad
    \squared{\sigma}{t} = \squared{s}{t} \int_0^t \frac{\squared{g}{\xi}}{\squared{s}{\xi}} \diff \xi ,
    $}
\end{equation}
and ${\gaussian(\,\cdot\,; \bmu, \bSigma)}$ denotes a multivariate Gaussian \cgls{pdf} with mean ${\bmu \inlinein \R^d}$ and covariance matrix ${\bSigma \inlinein \R^{d \times d}}$.
This allows to directly sample ${\bx_t}$ during training for any ${t}$ without having to numerically solve the forward \cgls{sde} in Eq.~\eqref{eq:linear_sde}.
However, during inference, since ${\bx_0}$ is unknown, noise cannot be sampled around ${s(t) \verythinspace \bx_0}$ to initialize the reverse process at ${t \inlineeq 1}$, unless ${s(1) \inlineeq 0}$.
In practice, it is common to use the zero-mean prior distribution ${\gaussian \big( \verythinspace \cdot \,; \mathbf{0}, \squared{\sigma}{1} \verythinspace \mathbf{I} \big)}$, even if this does not match ${p_{01}}$.

\subsection{Common SDEs}
\label{sec:common_sdes}

The joint parametrization of ${f(t)}$, ${g(t)}$, ${s(t)}$ and ${\sigma(t)}$ is commonly referred to as the \textit{noise schedule}.
One can first fix ${f(t)}$ and ${g(t)}$ and derive the corresponding ${s(t)}$ and ${\sigma(t)}$ from Eq.~\eqref{eq:marginals}, or alternatively one can first fix ${s(t)}$ and ${\sigma(t)}$ and derive the corresponding ${f(t)}$ and ${g(t)}$ after reverting Eq.~\eqref{eq:marginals},
\begin{equation}
    \label{eq:revert_marginals}
    f(t) = \frac{\diff}{\diff t} \Big[ \log s(t) \Big]
    \quad \text{and} \quad
    g(t) = s(t) \sqrt{\frac{\diff}{\diff t} \left[ \frac{\squared{\sigma}{t}}{\squared{s}{t}} \right]}.
\end{equation}

Recent studies~\cite{kingma2021variational, salimans2022progressive, hoogeboom2023simple, kingma2023understanding} argued that the noise schedule should be defined in terms of the log of the \cgls{snr} of the diffusion process ${\lambda(t)}$, %
\begin{equation}
    \label{eq:diffusion_snr}
    \lambda(t) = \log \frac{\squared{s}{t}}{\squared{\sigma}{t}}.
\end{equation}

Common choices for the drift and diffusion coefficients can be categorized into the \cgls{ve} and \cgls{vp} assumptions~\cite{song2021score}.

\subsubsection{\texorpdfstring{\cgls{ve}}{VE} assumption}

The \cgls{ve} assumption consists in setting ${f(t) \inlineeq 0}$, which leads to
\begin{equation}
    s(t) = 1
    \quad \text{and} \quad
    g(t) = \sqrt{\frac{\diff}{\diff t} \Big[ \squared{\sigma}{t} \Big]}.
\end{equation}
Song et al.~\cite{song2021score} proposed the following noise schedule,
\begin{equation}
    \squared{\sigma}{t} = \smin^2 \left[ \left( \frac{\smax}{\smin} \right)^{2t} - 1 \right],
\end{equation}
where ${\smin, \smax \inlinein \R^+, \smin \inlinelt \smax}$ are two hyperparameters.
As a consequence,
\begin{equation}
    \label{eq:ve_diffusion_coeff}
    g(t) = \smin \left( \frac{\smax}{\smin} \right)^t \sqrt{2 \log \frac{\smax}{\smin}}.
\end{equation}

\subsubsection{\texorpdfstring{\cgls{vp}}{VP} assumption}

The \cgls{vp} assumption consists in fixing ${\squared{s}{t} \inlineplus \squared{\sigma}{t} \inlineeq 1}$, which leads to
\begin{equation}
    \label{eq:vp_beta_coeff}
    f(t) = -\frac{1}{2} \beta(t)
    \quad \text{and} \quad
    g(t) = \sqrt{\beta(t)},
\end{equation}
where ${\beta : [0, \verythinnegativespace 1] \to \R^+}$ is a hyperparameter.
Song et al.~\cite{song2021score} proposed a linear noise schedule of the following form,
\begin{gather}
    \beta(t) = \bmin  + t \, (\bmax - \bmin),
\end{gather}
where ${\bmin, \bmax \inlinein \R^+, \bmin \inlinelt \bmax}$.
The perturbation kernel can then be derived using
\begin{equation}
    s(t) = e^{-\frac{1}{2} \int_0^t \beta(\xi) \diff \xi}
    \quad \text{and} \quad
    \squared{\sigma}{t} = 1 - e^{-\int_0^t \beta(\xi) \diff \xi}.
\end{equation}

Another common setting under the \cgls{vp} assumption is the family of cosine schedule~\cite{nichol2021improved, hoogeboom2023simple},
\begin{equation}
    \label{eq:cosine_schedule}
    \lambda(t) = -2 \log \left( \tan \frac{\pi t}{2} \right) + 2 \nu,
\end{equation}
where ${\nu}$ is a parameter controlling the center of the log-\cgls{snr} distribution.
The derivations of the perturbation kernel parameters ${s(t)}$ and ${\sigma(t)}$ and the \cgls{sde} coefficients ${f(t)}$ and ${g(t)}$ from ${\lambda(t)}$ are detailed in Appendix~\ref{app:cosine_schedule}.

\subsection{Elucidating the Design Space of Diffusion-Based Generative Models (EDM)}
\label{sec:edm}

Karras et al.~\cite{karras2022elucidating} showed that previous diffusion model formulations can be framed in a common framework denoted here as EDM.
More specifically, they showed that
\begin{itemize}
    \item The general \cgls{sde} in Eq.~\eqref{eq:linear_sde} is equivalent to a simpler, unscaled \cgls{sde} satisfied by the unscaled process ${\bxhat_t \inlineeq \frac{\bx_t}{s(t)}}$,
    \begin{equation}
        \diff \bxhat_t = \frac{g(t)}{s(t)} \diff \wiener.
    \end{equation}
    The perturbation kernel of the unscaled process is
    \begin{equation}
        \perturbkernelhat{\bxhat}{\bx} = \gaussian \big( \bxhat_t ; \bx_0, \squared{\sigmahat}{t} \verythinspace \mathbf{I} \big),
    \end{equation}
    where
    \begin{equation}
        \squared{\sigmahat}{t} = \frac{\squared{\sigma}{t}}{\squared{s}{t}} = \int_0^t \frac{\squared{g}{\xi}}{\squared{s}{\xi}} \diff \xi.
    \end{equation}
    Karras et al.~\cite{karras2022elucidating} suggested using ${s(t) \inlineeq 1}$.
    This is intuitive since the network is conditioned on the state index ${t}$ and can thus learn to undo the deterministic scaling ${s(t)}$ before denoising.
    This is in line with studies defining the noise schedule in terms of the log-\cgls{snr} ${\lambda(t)}$~\cite{kingma2021variational, salimans2022progressive, hoogeboom2023simple, kingma2023understanding}, as ${\squared{\sigmahat}{t} \inlineeq e^{-\lambda(t)}}$.
    The perturbation kernel of the scaled process ${\bx_t \inlineeq s(t) \verythinspace \bxhat_t}$ can be rewritten as
    \begin{equation}
        \perturbkernel{\bx} = \gaussian \big( \bx_t ; s(t) \verythinspace \bx_0, \squared{s}{t} \squared{\sigmahat}{t} \verythinspace \mathbf{I} \big).
    \end{equation}
    \item The marginal distribution ${p_t}$ can be written as
    \begin{equation}
        p_t(\bx_t) = s(t)^{-d} \phat ( \bxhat_t, \sigmahat(t) ),
    \end{equation}
    where ${\phat}$ is a mollified version of ${\pdata}$ obtained by adding i.i.d.\ Gaussian noise to the samples,
    \begin{equation}
        \phat(\cdot, \sigma) = \pdata \ast \gaussian(\mathbf{0}, \sigma^2 \verythinspace \mathbf{I}),
    \end{equation}
    where ${\ast}$ is the convolution operator between \cglspl{pdf}.
    As a consequence, the score function can be rewritten as
    \begin{equation}
        \label{eq:idpd_score}
        \score = \scorehat,
    \end{equation}
    which means that it can be evaluated on the unscaled process ${\bxhat_t}$.
    Let ${\idealdenoiser}$ denote the optimal denoiser that minimizes a weighted ${\mathcal{L}_2}$-loss between the clean training data ${\bx_0}$ and the output from the unscaled diffused data ${\bxhat_t}$,
    \begin{equation}
        \label{eq:denoiser_loss}
        \E_{\bx_0} \E_{t, \bxhat_t | \bx_0} \Big[ w(t) \left\| \idealdenoiser(\bxhat_t, t) - \bx_0 \right\|_2^2 \Big],
    \end{equation}
    where ${w : [0, \verythinnegativespace 1] \to \R^+}$ is the loss weighting function.
    Karras et al.~\cite{karras2022elucidating} showed that the score function can be expressed as a function of the optimal denoiser ${\idealdenoiser}$,
    \begin{equation}
        \scorehat = \frac{\idealdenoiser(\bxhat_t, t) - \bxhat_t}{s(t) \squared{\sigmahat}{t}}.
    \end{equation}
    \item The fitted denoiser ${\denoiser}$ trained to minimize Eq.~\eqref{eq:denoiser_loss} can be expressed as a function of the raw neural network layers ${\model}$.
    Such parametrization is referred to as \textit{preconditioning} and has the form
    \begin{equation}
        \label{eq:edm_precond}
        \resizebox{.81\hsize}{!}{$
        \displaystyle
        \denoiser(\bx, t) = \cskip(t) \bx + \cout(t) \model \big( \cin(t) \bx, \cnoise(t) \big).
        $}
    \end{equation}
    Karras et al.~\cite{karras2022elucidating} suggested the following parametrization for ${\cskip}$, ${\cout}$, ${\cin}$, ${\cnoise}$ and ${w}$ based on first principles (effective input unit variance, effective target unit variance and uniform effective loss weight),
    \begin{equation}
    \label{eq:preconditioning}
    \resizebox{.81\hsize}{!}{$
    \displaystyle
    \begin{gathered}
    \cskip(t) = \frac{\sdata^2}{\squared{\sigmahat}{t} + \sdata^2}, \quad
    \cout(t) = \frac{\sigmahat(t) \cdot \sdata}{\sqrt{\squared{\sigmahat}{t} + \sdata^2}},\\
    \cin(t) = \frac{1}{\sqrt{\squared{\sigmahat}{t} + \sdata^2}}, \quad
    \cnoise(t) = \frac{1}{4} \log \sigmahat(t),\\
    \text{and} \quad w(t) = \frac{\squared{\sigmahat}{t} + \sdata^2}{\squared{\sigmahat}{t} \cdot \sdata^2},
    \end{gathered}
    $}
    \end{equation}
    where ${\sdata^2}$ is the variance of the data distribution ${\pdata}$.
\end{itemize}

\subsection{Score-based Generative Model for Speech Enhancement (SGMSE)}
\label{sec:sgmse}

Welker et al.~\cite{welker2022speech} and Richter et al.~\cite{richter2023speech} adapted the diffusion model from Song et al.~\cite{song2021score} to perform speech enhancement by conditioning the clean speech generation process on the noisy speech signal.
They considered complex-valued \cgls{stft} representations of the signals flattened into ${d \inlineeq KT}$-dimensional vectors, where ${K}$ is the number of frequency bins and ${T}$ is the number of frames.
They trained the score model using pairs of clean and noisy speech signals ${\bx_0, \by \inlinein \C^d}$ in a supervised manner such that
\begin{equation}
    \scoremodel(\bx_t, \by, t) \simeq \scoreconditioned.
\end{equation}
They defined their diffusion equation as
\begin{equation}
    \label{eq:sgmse_sde}
    \resizebox{.89\hsize}{!}{$
    \displaystyle
    \diff \bx_t = -\gamma (\bx_t - \by) \diff t + \smin \left( \frac{\smax}{\smin} \right)^t \sqrt{2 \log \frac{\smax}{\smin}} \diff \wiener,
    $}
\end{equation}
that is ${\bff(t, \bx_t) \inlineeq -\gamma (\bx_t \inlineminus \by)}$ where ${\gamma}$ is a hyperparameter denoted as the \textit{stiffness}, and ${g(t)}$ is the same as the \cgls{ve} \cgls{sde} from~\cite{song2021score} and shown in Eq.~\eqref{eq:ve_diffusion_coeff}.
This \cgls{sde} was inspired by the \cgls{ou} \cgls{sde}~\cite{uhlenbeck1930theory} and results in an exponential decay of the mean of the diffusion process from ${\bx_0}$ towards ${\by}$.
We refer to this \cgls{sde} as the OUVE \cgls{sde}.

The drift coefficient ${\bff(t, \bx_t) \inlineeq -\gamma (\bx_t \inlineminus \by)}$ in Eq.~\eqref{eq:sgmse_sde} cannot be written in the form ${f(t) \verythinspace \bx_t}$, which means that the perturbation kernel cannot be calculated using Eqs.~\eqref{eq:perturb_ker} and \eqref{eq:marginals}.
Moreover, the model cannot be framed in the EDM framework, since we cannot find a scaling factor ${s(t)}$ such that the mean of the diffusion process is equal to ${s(t) \verythinspace \bx_0}$.
As a consequence, there does not seem to be an immediate way of applying the Heun-based sampler proposed in~\cite{karras2022elucidating} to integrate the reverse process.
Nonetheless, the perturbation kernel is Gaussian and can be calculated using Eqs.~(5.50) and (5.51) in~\cite{sarkka2019applied},
\begin{equation}
    \perturbkernelsgmse = \gaussian_\C \big( \bx_t ; \bmu(t), \squared{\sigma}{t} \verythinspace \mathbf{I} \big),
\end{equation}
where
\begin{gather}
    \bmu(t) = e^{-\gamma t} (\bx_0 - \by) + \by, \\
    \squared{\sigma}{t} = \frac{\smin^2}{1 + \gamma / \log \frac{\smax}{\smin}} \left[ \left( \frac{\smax}{\smin} \right)^{2t} - e^{-2 \gamma t} \right],
\end{gather}
and ${\gaussian_\C(\,\cdot\,; \bmu, \bSigma)}$ denotes a multivariate complex Gaussian \cgls{pdf} with mean ${\bmu \inlinein \C^d}$, covariance matrix ${\bSigma \inlinein \R^{d \times d}}$ and zero pseudo-covariance matrix.
The training loss in~\cite{welker2022speech,richter2023speech} was%
\footnote{\url{https://github.com/sp-uhh/sgmse/blob/c76f7ad04ec477ce0741387ec10e903dd3fac3d9/sgmse/model.py\#L115}}
\begin{equation}
    \label{eq:sgmse_loss}
    \E_{\bx_0, \by} \E_{t, \bz, \bx_t | (\bx_0, \by)} \Big[ \left\| \sigma(t) \scoremodel(\bx_t, \by, t) + \bz \right\|_2^2 \Big],
\end{equation}
where ${\bz \inlinesim \gaussian_\C(0, \mathbf{I})}$ and the score model was defined as%
\footnote{\url{https://github.com/sp-uhh/sgmse/blob/c76f7ad04ec477ce0741387ec10e903dd3fac3d9/sgmse/model.py\#L142}}%
\textsuperscript{,}%
\footnote{\url{https://github.com/sp-uhh/sgmse/blob/c76f7ad04ec477ce0741387ec10e903dd3fac3d9/sgmse/backbones/ncsnpp.py\#L413}}%
\textsuperscript{,}%
\footnote{\url{https://github.com/sp-uhh/sgmse/blob/c76f7ad04ec477ce0741387ec10e903dd3fac3d9/sgmse/backbones/ncsnpp.py\#L268}}%
\textsuperscript{,}%
\footnote{We note that compared to the implementation by Song et al.~\cite{song2021score}, the score model output was scaled using ${t}$ instead of ${\sigma(t)}$.
Moreover, the output convolution was placed after the output scaling, so the raw neural network output was actually ${-\model \big( \bx_t, \by, \log t \big)}$.
We believe this was unintended and we thus scale the output after the output convolution in our implementation.}
\begin{equation}
    \label{eq:sgmse_score_model}
    \scoremodel(\bx_t, \by, t) = - \frac{1}{t} \model \big( \bx_t, \by, \log t \big).
\end{equation}
To initialize the reverse process, noise was sampled around the noisy speech signal ${\by}$ instead of ${\bmu(1)}$, since the clean speech signal ${\bx_0}$ is unknown during inference.
There was thus a prior mismatch similar to Sect.~\ref{sec:diffusion_models}.
This was argued to be responsible for a drop in performance in speech enhancement~\cite{lay2023reducing}.

\section{Proposal to extend EDM to include SGMSE}
\label{sec:extending}

\subsection{Non-zero long-term mean SDE}

We propose to extend the EDM framework to account for a non-zero long-term mean ${\by}$ of the diffusion process.
Consider the following general \cgls{sde},
\begin{equation}
    \label{eq:non_linear_sde}
    \diff \bx_t = f(t) (\bx_t - \by) \diff t + g(t) \diff \wiener.
\end{equation}
This \cgls{sde} can be solved using the change of variable ${\bxtilde_t \inlineeq \bx_t \inlineminus \by}$ to obtain its perturbation kernel,
\begin{equation}
    \label{eq:perturb_kernel_extended}
    \perturbkernelsgmse = \gaussian \big( \bx_t ; s(t) (\bx_0 - \by) + \by, \squared{s}{t} \squared{\sigmahat}{t} \verythinspace \mathbf{I} \big),
\end{equation}
where
\begin{equation}
    \label{eq:non_linear_sde_marginals}
    s(t) = \exp \int_0^t f(\xi) \diff \xi
    \quad \text{and} \quad
    \squared{\sigmahat}{t} = \int_0^t \frac{\squared{g}{\xi}}{\squared{s}{\xi}} \diff \xi.
\end{equation}
We obtain the same scaling factor ${s(t)}$ as in Eq.~\eqref{eq:marginals}, but applied to the shifted process ${\bxtilde_t \inlineeq \bx_t \inlineminus \by}$.
For the OUVE \cgls{sde}, it can be identified that
\begin{gather}
    s(t) = e^{-\gamma t}, \label{eq:ouve_marginals_1} \\
    \squared{\sigmahat}{t} = \frac{\smin^2}{1 + \gamma / \log \frac{\smax}{\smin}} \left[ \left( e^{\gamma} \frac{\smax}{\smin} \right)^{2t} - 1 \right].
    \label{eq:ouve_marginals_2}
\end{gather}
Note that the OUVE \cgls{sde} does not fall under the VE assumption since ${f(t) \neq 0}$.

\subsection{Training loss and preconditioning}
\label{sec:preconditioning}

The loss function in Eq.~\eqref{eq:sgmse_loss} can be rewritten similarly to~\cite{karras2022elucidating} as a function of the denoiser ${\denoiser}$ to make the loss weighting and the preconditioning appear.
More specifically, it can be shown (see Appendix~\ref{app:sgmse_precond}) that the term inside the expectation operator in Eq.~\eqref{eq:sgmse_loss} is equal to
\begin{equation}
    \smallunderbrace{\frac{1}{\squared{\sigmahat}{t}}}_{w(t)} \left\| \denoiser(\bxtilde_0 + \unscalednoise, \by, t) - \bxtilde_0 \right\|_2^2, \label{eq:denoiser_loss_2}
\end{equation}
where ${\unscalednoise \inlinesim \gaussian_\C \big (0, \squared{\sigmahat}{t} \verythinspace \mathbf{I} \big)}$ and
\begin{equation}
    \label{eq:sgmse_precond}
    \resizebox{.89\hsize}{!}{$
    \displaystyle
    \denoiser(\bx, \by, t) = \\
    \!\!\! \smallunderbrace{1}_{\cskip} \!\!
    \bx \smallunderbrace{-\frac{s(t)\squared{\sigmahat}{t}}{t}}_{\cout}
    \model \big( \smallunderbrace{s(t)}_{\cin} \bx +
    \!\!\! \smallunderbrace{\by}_{\cshift} \!\!
    , \by, \smallunderbrace{\log t}_{\cnoise} \big).
    $}
\end{equation}
Compared to the EDM preconditioning in Eq.~\eqref{eq:edm_precond}, our preconditioning contains an additional parameter ${\cshift}$ to allow for including SGMSE.

\subsection{Sampling}

We cannot directly apply the Heun-based sampler from~\cite{karras2022elucidating} for our \cgls{sde} in Eq.~\eqref{eq:non_linear_sde} because Eq.~\eqref{eq:idpd_score} does not hold anymore.
However, a similar property can be derived.
Intuitively, we would like to evaluate the score function on the unshifted and unscaled process instead of the unscaled process.
More specifically, it can be shown (see Appendix~\ref{app:sampling_eq}) that
\begin{equation}
    \label{eq:sampling_eq}
    p_t(\bx_t | \by) = s(t)^{-d} \Big[ \ptildedata \ast \gaussian \big( \mathbf{0}, \squared{\sigmahat}{t} \verythinspace \mathbf{I} \big) \Big] \!\! \left( \frac{\bxtilde_t}{s(t)} \right),
\end{equation}
where ${\ptildedata : \bx \mapsto \pdata(\bx \inlineplus \by)}$ is a shifted version of ${\pdata}$ and ${\ast}$ is the convolution operator between \cglspl{pdf}.
Let $\bxtildehat_t \inlineeq \bxtilde_t/s(t) \inlineeq (\bx_t - \by)/s(t)$ denote the unshifted and unscaled process, and ${\ptildehat}$ the distribution between the brackets in Eq.~\eqref{eq:sampling_eq},
\begin{equation}
    \ptildehat(\cdot, \by, \sigma) = \ptildedata \ast \gaussian(\mathbf{0}, \sigma^2 \verythinspace \mathbf{I}).
\end{equation}
Therefore,
\begin{equation}
    \scoreconditioned = \scoretildehat.
\end{equation}
This equation is similar to Eq.~\eqref{eq:idpd_score} except that the diffusion process is now both unshifted and unscaled.
As a consequence, the Heun-based sampler from~\cite{karras2022elucidating} can be applied by evaluating the score model on the unshifted and unscaled process.

\subsection{Consequences}

\subsubsection{Alternative definition for the OUVE SDE and OUVP SDE}
\label{sec:alternative_ouve}

The OUVE \cgls{sde} is defined in~\cite{welker2022speech, richter2023speech} by setting the diffusion coefficient ${g(t)}$ to that of the \cgls{ve} \cgls{sde} in~\cite{song2021score}.
However, as pointed out in~\cite{karras2022elucidating}, defining ${g(t)}$ is of little use in practice as opposed to the perturbation kernel parameters ${s(t)}$ and ${\sigma(t)}$.
As a matter of fact, the resulting ${\sigma(t)}$ is different from that of the original \cgls{ve} \cgls{sde}.
From Eq.~\eqref{eq:ouve_marginals_1} it can be seen that the stiffness ${\gamma}$ is responsible for a simple scaling of the unshifted diffusion process by a factor ${s(t) \inlineeq e^{-\gamma t}}$.
With this in mind, a more practical definition for the OUVE \cgls{sde} in terms of the ${s(t)}$ and ${\sigmahat(t)}$ in Eq.~\eqref{eq:perturb_kernel_extended} can be formulated as follows,
\begin{equation}
    s(t) = e^{-\gamma t}
    \quad \text{and} \quad
    \squared{\sigmahat}{t} = \smin^2 \left[ \left( \frac{\smax}{\smin} \right)^{2t} - 1 \right].
\end{equation}
That is, the variance of the unscaled process is the same as the \cgls{ve} \cgls{sde}.
We refer to this \cgls{sde} as OUVE\textsubscript{2}.
The resulting drift and diffusion coefficients ${f(t)}$ and ${g(t)}$ in Eq.~\eqref{eq:non_linear_sde} are
\begin{equation}
    f(t) = -\gamma, \quad g(t) = e^{-\gamma t} \smin \! \left( \frac{\smax}{\smin} \right)^{t} \!\! \sqrt{2 \log \frac{\smax}{\smin}}.
\end{equation}

Similarly, a new \cgls{sde} termed the OUVP \cgls{sde} which combines the \cgls{ou} \cgls{sde} and the \cgls{vp} \cgls{sde} from~\cite{song2021score} can be defined by applying the additional scaling ${s(t) \inlineeq e^{-\gamma t}}$ to the already existing scaling from the \cgls{vp} \cgls{sde}.
Namely,
\begin{align}
    f(t) &= -\gamma - \frac{1}{2} \beta(t),
    &g(t) &= e^{-\gamma t} \sqrt{\beta(t)}, \\[3pt]
    s(t) &= e^{-\gamma t-\frac{1}{2} \int_0^t \beta(\xi) \diff \xi},
    &\squared{\sigmahat}{t} &= e^{\int_0^t \beta(\xi) \diff \xi} - 1.
\end{align}
Note that similar to the OUVE \cgls{sde} proposed in~\cite{welker2022speech,richter2023speech}, the OUVE\textsubscript{2} and OUVP \cglspl{sde} do not fall under the \cgls{ve} and \cgls{vp} assumptions.
In a recent work~\cite{guo2023variance}, an alternative \cgls{vp}-based formulation for speech enhancement is proposed, where the mean of the diffusion process is first interpolated between the clean speech signal ${\bx_0}$ and the noisy speech signal ${\by}$ before scaling.
However, the drift coefficient cannot be written in the form ${f(t) (\bx_t \inlineminus \by)}$ as in Eq.~\eqref{eq:non_linear_sde}, which prevents from framing the model in our extended EDM framework.

\subsubsection{The role of the drift term}
\label{sec:drift_role}

Previous studies on diffusion-based speech enhancement~\cite{welker2022speech, richter2023speech} have attributed their success to the addition of the stiffness term ${\gamma}$ to provoke a progressive drift of the mean of the diffusion process towards the noisy speech ${\by}$.
This was argued to enable the system to reconstruct speech corrupted by environmental noises that differ from the stationary Gaussian noise of the diffusion process~\cite{richter2023speech}.
However, as seen in Eq.~\eqref{eq:ouve_marginals_1}, the addition of the stiffness term ${\gamma}$ is only responsible for a deterministic scaling of the unshifted diffusion process by a factor ${e^{-\gamma t}}$.
This means the need for ${\gamma}$ is in contradiction with the suggested ${s(t) \inlineeq 1}$ in~\cite{karras2022elucidating} and the intuition that the network can learn to undo this scaling.
In our view, the process should be seen as fully generative with ${\by}$ only serving as a conditioner, without the need for a progressive transformation between the distributions of ${\bx_0}$ and ${\by}$.
The noisy speech ${\by}$ should be treated similarly to a text prompt in image generation, the only difference being that in speech enhancement, the conditioner has the same modality as the output.

\section{Experimental setup}
\label{sec:experimental}

\subsection{Datasets}

We consider two noisy speech datasets, namely VBDMD~\cite{valentini2016speech} and our own MultiCorpus dataset.
The speech enhancement systems are evaluated in matched conditions using the training, validation and test splits from the same dataset (no cross-dataset evaluation).

\subsubsection{VBDMD}
The VoiceBank+DEMAND (VBDMD) 28-speaker dataset~\cite{valentini2016speech} is a publicly available and popular benchmark dataset for single-channel speech enhancement.
The noisy speech mixtures were generated by mixing clean speech utterances from the Voice Bank corpus~\cite{veaux2013voice} with 10 noise types from the DEMAND database~\cite{thiemann2013demand} at SNRs of \SIlist{0;5;10;15}{\decibel}.
The test mixtures were obtained using two different speakers and five different noise types at SNRs of \SIlist{2.5;7.5;12.5;17.5}{\decibel}.
We remove speakers p226 and p287 from the training set to use them for validation similarly to~\cite{lu2021study,lu2022conditional,welker2022speech,richter2023speech}.
The duration of the training, validation and test sets is \SI{8}{\hour}\,\SI{45}{\minute}, \SI{37}{\minute} and \SI{34}{\minute} respectively.

\subsubsection{MultiCorpus}
\label{sec:multi_corpus}

We generate a diverse and acoustically adverse MultiCorpus dataset by mixing reverberant speech and reverberant noise using speech utterances, noise segments and \cglspl{brir} from an ensemble of five different speech corpora, five noise databases and five \cgls{brir} databases.
The selected databases are the same as in~\cite{gonzalez2023assessing}.
Namely for the speech, we use TIMIT~\cite{garofolo1993timit}, LibriSpeech (100-hour version)~\cite{panayotov2015librispeech}, WSJ SI-84~\cite{paul1992design}, Clarity~\cite{cox2022clarity} and VCTK~\cite{veaux2013voice}.
For the noise, we use TAU~\cite{heittola2019tau}, NOISEX~\cite{varga1993noisex}, ICRA~\cite{dreschler2001icra}, DEMAND~\cite{thiemann2013demand} and ARTE~\cite{weisser2019ambisonic}.
Finally for the \cglspl{brir}, we use Surrey~\cite{hummersone2010surrey}, ASH~\cite{shanon2021ash}, BRAS~\cite{brinkmann2021bras}, CATT~\cite{catt_brirs} and AVIL~\cite{marschall2017database}.
Each mixture is simulated by convolving one clean speech utterance and up to three noise segments with \cglspl{brir} from different and random spatial locations between \SI{-90}{\degree} and \SI{90}{\degree} in the same room.
The convolution outputs are then mixed at a random \cgls{snr} uniformly distributed between \SI{-5}{\decibel} and \SI{10}{\decibel}.
The \cgls{snr} is defined as the energy ratio between the direct-sound part of the speech signal and the background signal, which consists of the direct-sound and reverberant parts of the noise signals as well as the reverberant part of the speech signal.
The direct-sound part of the speech signal includes early reflections up to a boundary of \SI{50}{\milli\second}, which was shown to be beneficial for speech intelligibility~\cite{roman2013speech}.
The different signals are then created by splitting the \cglspl{brir} into a direct-sound and a reverberant component using a windowing procedure described in~\cite{zahorik2002direct}, and averaged across left and right channels.
For the speech, \SI{80}{\percent} of the utterances in each corpus is reserved for training and \SI{20}{\percent} for testing.
For the noise, \SI{80}{\percent} of each recording is reserved for training and \SI{20}{\percent} for testing.
For the \cglspl{brir}, every second \cgls{brir} from each room is reserved for training and the rest for testing.
A validation set is created using the same split as the training set, but with different random mixture realizations.
The duration of the training, validation and test sets is \SI{10}{\hour}, \SI{30}{\minute} and \SI{30}{\minute} respectively.

\subsection{Data pre-processing}
\label{sec:pre_proc}

We use the same data representation as in~\cite{welker2022speech,richter2023speech}.
The clean and noisy speech signals are sampled at \SI{16}{\kilo\hertz}.
The \cgls{stft} is calculated using a frame length of 512 samples, a hop size of 128 samples and a Hann window.
We discard the Nyquist component to obtain ${K \inlineeq 256}$ frequency bins.
An amplitude transformation is then applied to each \cgls{stft} coefficient ${c \inlinein \C}$ to compensate for their heavy-tailed distribution,
\begin{equation}
    \label{eq:stft_transform}
    \tilde c = A |c|^\alpha e^{i \angle c},
\end{equation}
where ${\tilde c}$ is the transformed coefficient, ${A \inlinein \R^+}$ is a scaling factor and ${\alpha \inlinein \R^+}$ is a compression factor.
We use the same values for ${A}$ and ${\alpha}$ as in~\cite{welker2022speech,richter2023speech}, namely ${A \inlineeq 0.15}$ and ${\alpha \inlineeq 0.5}$.

\subsection{Neural network}
\label{sec:neural_network}

We use the same NCSN++M neural network architecture as in~\cite{lemercier2023analysing,lemercier2023storm}.
This is a scaled-down version of the NCSN++ neural network initially proposed by Song et al.~\cite{song2021score} and used in~\cite{welker2022speech,richter2023speech}.
It consists of a multi-resolution U-Net~\cite{ronneberger2015unet}-like architecture with an additional parallel progressive growing path.
Compared to~\cite{welker2022speech,richter2023speech}, the number of layers in the encoder/decoder was reduced to four, the number of residual blocks in each layer was reduced to one, and all attention blocks were removed except in the bottleneck layer.
This reduces the number of parameters from \SI{65}{\mega\nothing} to \SI{27.8}{\mega\nothing} without losing performance~\cite{lemercier2023analysing,lemercier2023storm}.

\subsection{Training}
\label{sec:training}

The neural network is trained for 300 epochs using the Adam optimizer~\cite{kingma2015adam} and a learning rate of ${1e^{-4}}$.
Due to the varying duration of the training sequences, the complex-valued spectrograms in ${\C^{K \times T}}$ with variable number of frames ${T}$ are batched using a bucketing strategy~\cite{gonzalez2023batching} with 10 buckets and a dynamic batch size of \SI{32}{\second}.
Exponential moving average of the neural network weights is applied with a decay rate of 0.999, as this was reported to greatly improve results for image generation~\cite{song2020improved,song2021score}.
We randomly sample ${t \inlinesim \uniform{t_\epsilon}{1}}$ with ${t_\epsilon \inlineeq 0.01}$ during training to avoid instabilities for very small ${t}$ values similarly to previous studies~\cite{song2021score,richter2023speech,guo2023variance}.

\subsection{Baselines}
\label{sec:baselines}

We use Conv-TasNet~\cite{luo2019conv}, DCCRN~\cite{hu2020dccrn}, MANNER~\cite{park2022manner} and SGMSE+M~\cite{lemercier2023analysing} as baselines.
They are all trained on full-length mixtures for 100 epochs using the Adam optimizer~\cite{kingma2015adam} and a bucketing strategy~\cite{gonzalez2023batching} with 10 buckets and a dynamic batch size of \SI{128}{\second}, except SGMSE+M which is trained for 300 epochs with a dynamic batch size of \SI{32}{\second} as in Sect.~\ref{sec:training}.

\subsubsection{Conv-TasNet}

An end-to-end fully convolutional network designed for single-channel multi-speaker speech separation.
It can be applied to speech enhancement as described in~\cite{koyama2020exploring,kinoshita2020improving} by setting the number of separated sources to ${K \inlineeq 1}$.
We use a learning rate of ${1e^{-3}}$ and the \cgls{snr} loss instead of the \cgls{sisnr}~\cite{leroux2019sdr} loss to avoid scaling the output signal.
This was shown to provide superior performance in terms of multiple perceptual metrics~\cite{koyama2020exploring}.
It has \SI{4.9}{\mega\nothing} parameters.

\subsubsection{DCCRN}

A causal U-Net~\cite{ronneberger2015unet}-based network operating in the \cgls{stft} domain.
It leverages complex-valued convolutions~\cite{trabelsi2018deep} in the encoder/decoder and a long short-term memory network~\cite{hochreiter1997long} in the bottleneck layer to predict a complex-valued mask.
We train it with the \cgls{snr} loss instead of the \cgls{sisnr} loss similar to Conv-TasNet.
We set the total number of encoder channels to ${[32, \verythinnegativespace 64, \verythinnegativespace 128, \verythinnegativespace 256, \verythinnegativespace 256, \verythinnegativespace 256]}$ and use the ``E'' enhancement strategy.
We use regular batch normalization~\cite{ioffe2015batch} along the real and imaginary axes separately instead of complex batch normalization~\cite{trabelsi2018deep} as this improves training time and stability without reducing performance.
This is in line with recent evidence that the performance of DCCRN cannot be attributed to complex-valued operations~\cite{wu2023rethinking}.
We use a learning rate of ${1e^{-4}}$ and gradient clipping with a maximum norm of 5 to further improve stability.
It has \SI{3.7}{\mega\nothing} parameters.

\subsubsection{MANNER}

A multi-view attention network combining channel attention~\cite{woo2018cbam} with local and global attention along two signal scales similar to dual-path models~\cite{luo2020dual}.
It is trained with a combination of a time-domain ${\mathcal{L}_1}$-loss and a multi-resolution \cgls{stft} loss~\cite{yamamoto2020parallel} as in~\cite{defossez2020real}.
We use the ``small'' version provided in~\cite{park2022manner}, which presents multi-attention blocks in the last encoder/decoder layers only, as this greatly reduces computational cost without significantly reducing performance~\cite{park2022manner}.
We use the OneCycleLR scheduler~\cite{smith2019super} with a minimum and maximum learning rate of ${1e^{-5}}$ and ${1e^{-2}}$ respectively as in~\cite{park2022manner}.
It has \SI{21.2}{\mega\nothing} parameters.

\subsubsection{SGMSE+M}

A diffusion-based system with the same neural network backbone as described in Sect.~\ref{sec:neural_network}, but using the original preconditioning, OUVE \cgls{sde} and \cgls{pc} sampler as in~\cite{welker2022speech,richter2023speech}.
The preconditioning corresponds to Eq.~\eqref{eq:sgmse_precond} and the \cgls{sde} corresponds to Eq.~\eqref{eq:sgmse_sde}.
We use ${\gamma \inlineeq 1.5}$ and one step of annealed Langevin dynamics correction with size ${r \inlineeq 0.5}$ as suggested in~\cite{richter2023speech}.
The data pre-processing and neural network training match Sect.~\ref{sec:pre_proc} and Sect.~\ref{sec:training} respectively.
By comparing with this baseline, we are able to assess the influence of the preconditioning, the \cgls{sde} and the solver.
It has \SI{27.8}{\mega\nothing} parameters.

\subsection{Objective metrics}

We evaluate the performance of the systems in terms of \cgls{pesq}~\cite{recommendation2001perceptual}, \cgls{estoi}~\cite{jensen2016algorithm} and \cgls{snr}.
The results are reported in terms of average metric improvement between the unprocessed input mixture and the enhanced output signal.
The improvements are denoted as ${\dpesq}$, ${\destoi}$ and ${\dsnr}$ respectively.
Note that for the MultiCorpus dataset, the reference signal of each metric is defined as the direct-sound component of the speech signal averaged across left and right channels (see Sect.~\ref{sec:multi_corpus} for details).

\subsection{Experiments}

We investigate three design aspects of diffusion-based speech enhancement in light of the extended EDM framework proposed in Sect.~\ref{sec:extending}: the preconditioning and loss weighting, the \cgls{sde}, and the amount of stochasticity injected when integrating the reverse process.
Note that these design aspects do not affect the computational complexity of the system, since the neural network architecture and the training procedure are the same.
For each experiment, the reverse diffusion process is integrated by uniformly discretizing the time axis between ${t \inlineeq 1}$ and ${t \inlineeq 0}$.
The number of discretization steps ${n_{\mathrm{steps}}}$ is varied in powers of 2 between 1 and 64.
To initialize the reverse process, we sample from ${\gaussian_\C \big( \verythinspace \cdot \,; \by, \squared{\sigma}{1} \verythinspace \mathbf{I} \big)}$ as in~\cite{welker2022speech,richter2023speech}.

\subsubsection{Preconditioning and loss weighting}

Eq.~\eqref{eq:sgmse_precond} shows the preconditioning and loss weighting corresponding to~\cite{welker2022speech,richter2023speech}.
However, this preconditioning is different from Eq.~\eqref{eq:preconditioning}, which was derived from first principles and suggested in~\cite{karras2022elucidating}.
We thus investigate the effect of each preconditioning parameter ${\cskip}$, ${\cout}$, ${\cin}$, ${\cnoise}$ and ${\cshift}$ as well as the loss weighing ${w}$.
To do so, we incrementally change the preconditioning parameters and loss weighting from the values used in~\cite{welker2022speech,richter2023speech} and shown in Eq.~\eqref{eq:sgmse_precond} to the values suggested in~\cite{karras2022elucidating} and shown in Eq.~\eqref{eq:preconditioning}.
For ${\cshift}$, since this parameter was introduced by us in Sect.~\ref{sec:preconditioning} to include SGMSE in EDM, there is no recommended value in~\cite{karras2022elucidating}.
Since we see no theoretical foundation to set ${\cshift}$ to ${\by}$ as in~\cite{welker2022speech,richter2023speech} and Eq.~\eqref{eq:sgmse_precond}, we set ${\cshift}$ to zero when investigating its effect.
For the parameter ${\sdata}$, the value used in~\cite{karras2022elucidating} was 0.5, as this corresponds to the variance of pixel values in popular image datasets when centered around ${[-1, \verythinnegativespace 1]}$.
However in our case, the variance of complex \cgls{stft} coefficients after applying the transform in Eq.~\eqref{eq:stft_transform} is around 0.1.
We thus set ${\sdata \inlineeq 0.1}$.
We use the same OUVE \cgls{sde} and \cgls{pc} sampler as in~\cite{welker2022speech,richter2023speech}.

\subsubsection{SDE}

\begin{figure*}[t]
    \centering
    \myabovedoublefigureskip
    \subfloat[Speech enhancement case]{\includegraphics[width=.49\linewidth]{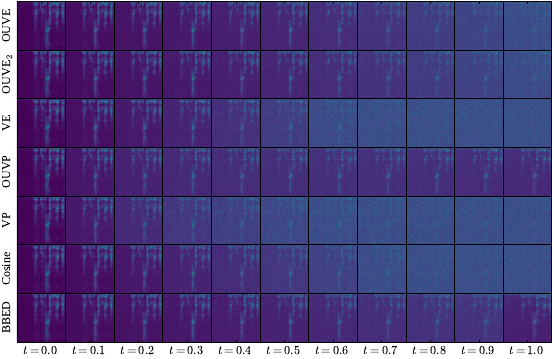}}
    \hfil
    \subfloat[Image generation case]{\includegraphics[width=.49\linewidth]{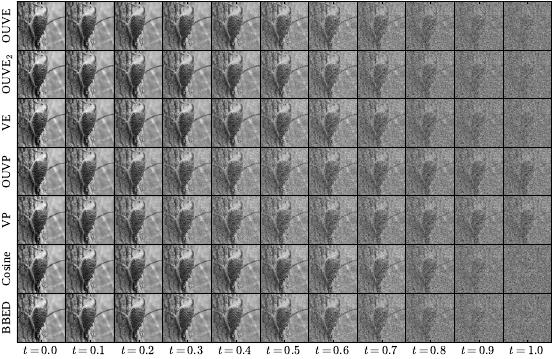}}
    \caption{Example evolution of the diffusion process for each \cgls{sde} investigated.
    In the speech enhancement case (a), the initial state of the diffusion process is a complex \cgls{stft} representation of the clean speech ${\bx_0}$.
    Each \cgls{sde} provokes a drift of the mean of the process towards the noisy speech ${\by}$ at a different rate.
    The magnitude of the complex-valued diffusion process is plotted.
    In the image generation case (b), the initial state is a clean image and no drift towards the conditioner occurs, which is equivalent to setting ${\by \inlineeq \mathbf{0}}$ in our framework.}
    \mybelowcaptionskip
    \label{fig:checkboards}
\end{figure*}

\begin{figure}
    \centering
    \myabovesinglefigureskip
    \includegraphics{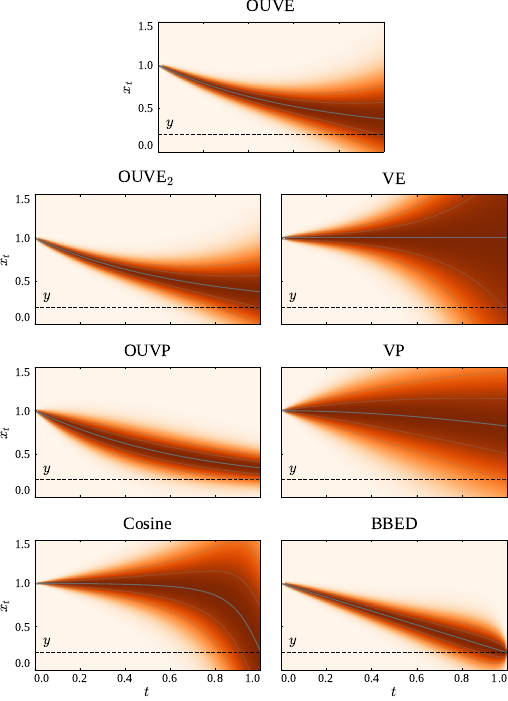}
    \myabovecaptionskip
    \caption{Illustration of the different \cglspl{sde} investigated.}
    \mybelowcaptionskip
    \label{fig:sdes}
\end{figure}

We fix the preconditioning parameters and the loss weighting to the values shown in Eq.~\eqref{eq:preconditioning} as in~\cite{karras2022elucidating}, since these are motivated by first principles.
We then experiment with the following \cglspl{sde}:
\begin{itemize}
    \item \textit{OUVE}: Proposed in~\cite{welker2022speech,richter2023speech} and shown in Eq.~\eqref{eq:sgmse_sde}.
    It allows for a progressive drift of the mean of the diffusion process towards the noisy speech ${\by}$.
    We use ${\sigma_{\min} \inlineeq 0.05}$, ${\sigma_{\max} \inlineeq 0.5}$ and ${\gamma \inlineeq 1.5}$ as in~\cite{welker2022speech,richter2023speech}.
    \item \textit{OUVE\textsubscript{2}}: Alternative definition for the OUVE \cgls{sde} proposed in Sect.~\ref{sec:alternative_ouve}, which results in the same unscaled variance ${\squared{\sigmahat}{t}}$ as the \cgls{ve} \cgls{sde} from~\cite{song2021score}.
    We set ${\sigma_{\min} \inlineeq 0.04}$, ${\sigma_{\max} \inlineeq 1.7}$ and ${\gamma \inlineeq 1.5}$ empirically based on preliminary experiments.
    \item \textit{VE}: Original \cgls{ve} \cgls{sde} from image generation literature~\cite{song2021score} and described in Sect.~\ref{sec:common_sdes}.
    Similar to OUVE\textsubscript{2} but sets ${s(t) \inlineeq 1}$ (i.e.\ ${\gamma \inlineeq 0}$), meaning no drift of the mean of the diffusion process occurs.
    To the best of our knowledge, this was not used for speech enhancement before.
    We set ${\sigma_{\min} \inlineeq 0.04}$ and ${\sigma_{\max} \inlineeq 1.7}$ as for OUVE\textsubscript{2}.
    \item \textit{OUVP}: Combination of the \cgls{ou} and \cgls{vp} \cglspl{sde} proposed in Sect.~\ref{sec:alternative_ouve}.
    We set ${\beta_{\min} \inlineeq 0.01}$, ${\beta_{\max} \inlineeq 1}$ and ${\gamma \inlineeq 1.5}$ empirically based on preliminary experiments.
    \item \textit{VP}: Original \cgls{vp} \cgls{sde} from image generation literature~\cite{song2021score} and described in Sect.~\ref{sec:common_sdes}.
    Similar to OUVP but without the additional drift of the mean of the diffusion process (i.e.\ ${\gamma \inlineeq 0}$).
    To the best of our knowledge, this was not used for speech enhancement before.
    We set ${\beta_{\min} \inlineeq 0.01}$, ${\beta_{\max} \inlineeq 1}$ and ${\gamma \inlineeq 1.5}$ as for OUVP.
    \item \textit{Cosine}: Shifted-cosine noise schedule from image generation~\cite{hoogeboom2023simple} and shown in Eq.~\eqref{eq:cosine_schedule}.
    To the best of our knowledge, this was not used for speech enhancement before, except in our previous study~\cite{gonzalez2024diffusion}.
    We clamp ${\lambda(t) \inlinegeq \lambda_{\min}}$ and ${\beta(t) \inlineleq \beta_{\max}}$ to prevent instabilities around ${t \inlineeq 1}$.
    We set ${\nu \inlineeq 1.5}$, ${\lambda_{\min} \inlineeq -12}$ and ${\beta_{\max} \inlineeq 10}$ empirically based on preliminary experiments.
    \item \textit{BBED}: Brownian Bridge with Exponential Diffusion Coefficient (BBED) proposed in~\cite{lay2023reducing}.
    It replaces the exponential drift from the OUVE \cgls{sde} with a linear interpolation between the clean speech ${\bx_0}$ and the noisy speech ${\by}$.
    This reduces the mismatch between the mean of the diffusion process at ${t \inlineeq 1}$ and the noisy speech ${y}$ around which noise is sampled to initialize the reverse process.
    This was argued to improve speech enhancement performance compared to the OUVE \cgls{sde}~\cite{lay2023reducing}.
    We set ${c \inlineeq 0.01}$ and ${k \inlineeq 10}$ as in~\cite{lay2023reducing} and use the change of variable ${t \gets t_{\max} \verythinnegativespace \cdot \verythinnegativespace t}$ with ${t_{\max} \inlineeq 0.999}$ to prevent instabilities around ${t \inlineeq 1}$.
\end{itemize}
The hyper-parameters for each \cgls{sde} significantly affect the performance. For each \cgls{sde}, either preliminary experiments are conducted to find the best hyper-parameters, or the values suggested in previous studies are used. The preliminary experiments are not detailed here for brevity.

The different \cglspl{sde} are illustrated in the one-dimensional case in Fig.~\ref{fig:sdes}, where the different drifts between the clean speech ${\bx_0}$ and the noisy speech ${\by}$ can be observed.
It can be seen that the \cgls{ve} \cgls{sde} shows no drift towards the noisy speech ${\by}$ and the perturbation kernel stays centered around the clean speech ${\bx_0}$.
The \cgls{vp} \cgls{sde} shows a very slight drift compared to the OUVP \cgls{sde}.
The cosine and BBED \cglspl{sde} show a very small mismatch between the mean of the diffusion process at ${t \inlineeq 1}$ and the noisy speech ${\by}$.
Note that the reverse process is still initialized by sampling from a Gaussian centered around ${\by}$, even when using the \cgls{ve} and \cgls{vp} \cglspl{sde}.
This means that for the \cgls{ve} and \cgls{vp} \cglspl{sde}, the prior mismatch is substantially larger compared to the other \cglspl{sde}.

Figure~\ref{fig:checkboards} shows an example of the evolution of the diffusion process for each \cgls{sde}, in the case of speech enhancement where the long-term mean of the diffusion process is the noisy speech ${\by}$, and in the case of image generation where the long-term mean of the diffusion process is zero.
In the image generation case, the OUVE\textsubscript{2} and \cgls{ve} \cglspl{sde} are equivalent, since no drift towards the conditioner occurs.
This is modeled by setting ${\by \inlineeq \mathbf{0}}$ in our framework.
However, in the speech enhancement case, the last state of the OUVE\textsubscript{2} process is the noisy speech ${\by}$ with a small amount of Gaussian noise added, while the \cgls{ve} process is the clean speech ${\bx_0}$ with a larger amount of Gaussian noise added.
Similar observations can be made for the OUVP and \cgls{vp} \cglspl{sde}.
Finally, while the cosine and BBED perturbation kernels are both centered around the noisy speech ${\by}$ at ${t \inlineeq 1}$, the variance is very high for the cosine \cgls{sde}, while it is very small for the BBED \cgls{sde}.

\subsubsection{Stochasticity injected in reverse process}
\label{sec:sampler_exp}

We fix the preconditioning parameters and the loss weighting as in the previous experiment.
We also chose the cosine \cgls{sde}, since it was shown to provide superior performance in image generation compared to linear noise schedules~\cite{nichol2021improved,hoogeboom2023simple,kingma2023understanding}.
To integrate the reverse process, we use either the \cgls{pc} sampler from Song et al.~\cite{song2021score}, which has been extensively used for speech enhancement~\cite{welker2022speech,richter2023speech,lemercier2023analysing,lay2023reducing,lemercier2023storm}, or the 2\textsuperscript{nd} order Heun-based sampler from the EDM framework~\cite{karras2022elucidating}.
Pseudo-code for both samplers is provided in Alg.~\ref{alg:pc_sampler} and Alg.~\ref{alg:edm_sampler}.
In the case of the \cgls{pc} sampler, the parameter controlling the amount of stochasticity injected in the reverse process is the step size of the annealed Langevin dynamics correction ${r}$.
In the case of the Heun-based sampler, this is controlled by the parameter ${S_\mathrm{churn}}$.
We vary ${r \inlinein [0, \verythinnegativespace 1]}$ when using the \cgls{pc} sampler and ${S_\mathrm{churn} \inlinein [0, \verythinnegativespace 20] \verythinnegativespace \cup \verythinnegativespace \{ \infty \}}$ when using the Heun-based sampler.
Note that the Heun-based sampler algorithm clamps the amount of noise injected in the reverse process such that it does not exceed the amount of noise already present in the current diffusion state.
The clamping threshold increases with the number of sampling steps ${n_{\mathrm{steps}}}$.
For the largest number of sampling steps ${n_{\mathrm{steps}} \inlineeq 64}$ considered in this study, the threshold is ${S_\mathrm{churn} \inlineeq 26.5}$, which is why we do not investigate ${S_\mathrm{churn}}$ values between 20 and ${\infty}$.
The remaining parameters for the Heun-based sampler are empirically set to ${S_{\mathrm{noise}} \inlineeq 1}$, ${S_{\min} \inlineeq 0}$ and ${S_{\max} \inlineeq \infty}$ based on preliminary experiments.
To the best of our knowledge, the Heun-based sampler has not been used for speech enhancement before, except in our previous study~\cite{gonzalez2024diffusion} where ${S_\mathrm{churn}}$ was fixed to ${\infty}$.
For the \cgls{pc} sampler, we use a single corrector step as suggested in~\cite{richter2023speech}.
This also ensures that the number of neural network evaluations is the same with both samplers for a fixed ${n_{\mathrm{steps}}}$.
Note that while the effect of the step size ${r}$ on speech enhancement performance was investigated in~\cite{richter2023speech}, this was done with a fixed ${n_{\mathrm{steps}}}$.
It is thus unclear if the optimal step size ${r}$ is the same for different ${n_{\mathrm{steps}}}$ values.
Moreover, this was investigated in terms of \cgls{pesq} and \cgls{sisnr}, but not in terms of \cgls{estoi}.

\begin{algorithm}[t]
\small
\caption{PC sampler~\cite{song2021score}}
\label{alg:pc_sampler}
\begin{algorithmic}
\Require ${1\!=\!t_0\!>\!t_1\!>\!\ldots\!>\!t_{n_\mathrm{steps}}\!=\!0}$ the sampling step boundaries
\State \textbf{sample} $\bx \sim \gaussian_\C (\mathbf{y}, s(t_0) \, \sigmahat^2(t_0) \, \mathbf{I})$
\For{$i \gets 0, \ldots, n_\mathrm{steps} - 1$}
    \State \tikzmk{A} \textbf{sample} $\mathbf{z} \sim \gaussian_\C (\mathbf{0}, \mathbf{I})$
    \Comment Corrector step
    \State $\varepsilon \gets 2 \left( r \, s(t_i) \, \sigmahat(t_i) \right)^2$
    \State $\bxtildehat \gets \frac{\bx - \by}{s(t_i)}$
    \State $\mathbf{s} \gets \frac{\denoiser(\bxtildehat, \by, t_i) - \bxtildehat}{s(t_i) \sigmahat^2(t_i)}$
    \State \tikzmk{B} $\bx \gets \bx + \varepsilon \, \mathbf{s} + \sqrt{2 \varepsilon} \, \mathbf{z}$
    \boxit{red!40}{0.935\linewidth}{0pt}{1pt}

    \State \tikzmk{A} $\bxtildehat \gets \frac{\bx - \by}{s(t_i)}$ \Comment Predictor step
    \State $\mathbf{s} \gets \frac{\denoiser(\bxtildehat, \by, t_i) - \bxtildehat}{s(t_i) \sigmahat^2(t_i)}$
    \If{$i < n_\mathrm{steps} - 1$}
        \State \textbf{sample} $\mathbf{z} \sim \gaussian_\C (\mathbf{0}, \mathbf{I})$
        \State $\mathbf{d} \gets f(t_i)(\bx - \by) - g^2(t_i) \, \mathbf{s}$
        \State $\bx \gets \bx + (t_{i+1} - t_i) \, \mathbf{d} + g(t_i) \sqrt{t_{i} - t_{i+1}} \, \mathbf{z}$
    \Else
        \State $\mathbf{d} \gets f(t_i)(\bx - \by) - \frac{1}{2} g^2(t_i) \, \mathbf{s}$
        \State \tikzmk{B} $\bx \gets \bx + (t_{i+1} - t_i) \, \mathbf{d}$
        \boxit{cyan!40}{0.88\linewidth}{0pt}{1pt}
    \EndIf
\EndFor
\State \textbf{return} $\bx$
\end{algorithmic}
\end{algorithm}

\begin{algorithm}[t]
\small
\caption{Heun-based sampler~\cite{karras2022elucidating}}
\label{alg:edm_sampler}
\begin{algorithmic}
\Require ${1\!=\!t_0\!>\!t_1\!>\!\ldots\!>\!t_{n_\mathrm{steps}}\!=\!0}$ the sampling step boundaries
\State \textbf{sample} $\bx \sim \gaussian_\C (\mathbf{y}, s(t_0) \, \sigmahat^2(t_0) \, \mathbf{I})$
\For{$i \gets 0, \ldots, n_\mathrm{steps} - 1$}
    \State \tikzmk{A} \textbf{sample} $\mathbf{z} \sim \gaussian_\C (\mathbf{0}, S_\mathrm{noise}^2 \mathbf{I})$ \Comment Stochastic step
    \State $\gamma \gets \begin{cases}
        \min\!\big( \frac{S_\mathrm{churn}}{N}, \sqrt{2} - 1 \big) & \text{if } \sigmahat(t_i) \in [S_\mathrm{min}, S_\mathrm{max}] \\[-1pt]
        0 & \text{otherwise}
    \end{cases}$
    \State $\sigmahat' \gets (1 + \gamma) \, \sigmahat(t_i)$
    \State $t' \gets \sigmahat^{-1}(\sigmahat')$
    \State \tikzmk{B} $\bx' \gets \frac{s(t')}{s(t_i)}(\bx - \by) + \by + s(t')\sqrt{\sigma'^2 - \sigmahat^2(t_i)} \, \mathbf{z}$
    \boxit{red!40}{0.935\linewidth}{-2pt}{1pt}

    \State \tikzmk{A} $\bxtildehat \gets \frac{\bx' - \by}{s(t')}$ \Comment Deterministic step
    \State $\mathbf{s} \gets \frac{\denoiser(\bxtildehat, \by, t') - \bxtildehat}{s(t') \sigmahat'^2}$
    \State $\mathbf{d} \gets f(t')(\bx' - \by) - \frac{1}{2} g^2(t') \, \mathbf{s}$
    \State $\bx \gets \bx' + (t_{i+1} - t') \, \mathbf{d}$

    \If{$i < n_\mathrm{steps} - 1$}
        \State $\bxtildehat \gets \frac{\bx - \by}{s(t_{i+1})}$
        \State $\mathbf{s} \gets \frac{\denoiser(\bxtildehat, \by, t_{i+1}) - \bxtildehat}{s(t_{i+1}) \sigmahat(t_{i+1})}$
        \State $\mathbf{d}' \gets f(t_{i+1})(\bx - \by) - \frac{1}{2} g^2(t_{i+1}) \, \mathbf{s}$
        \State \tikzmk{B} $\bx \gets \bx' + \frac{1}{2} (t_{i+1} - t') (\mathbf{d} + \mathbf{d}')$
        \boxit{cyan!40}{0.88\linewidth}{-2.5pt}{2.5pt}
    \EndIf
\EndFor
\State \textbf{return} $\bx$
\end{algorithmic}
\end{algorithm}

\section{Results}
\label{sec:results}

\subsection{Preconditioning and loss weighting}

\begin{figure}
    \centering
    \myabovesinglefigureskip
    \includegraphics{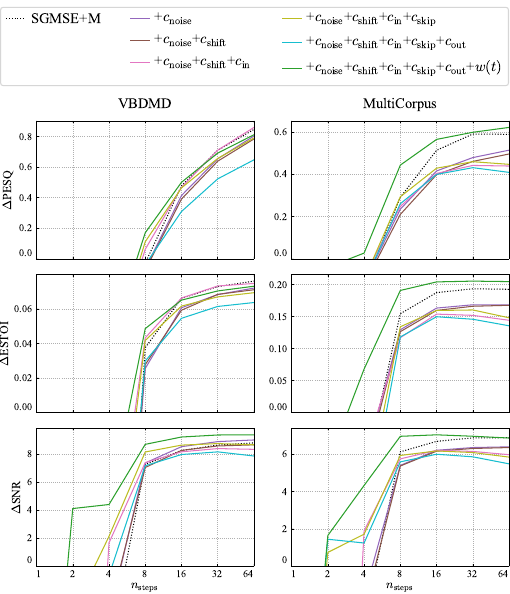}
    \myabovecaptionskip
    \caption{Speech enhancement performance as a function of the number of sampling steps ${n_{\mathrm{steps}}}$ when changing the preconditioning parameters incrementally from the values in Eq.~\eqref{eq:sgmse_precond} used in~\cite{welker2022speech,richter2023speech} to the values in Eq.~\eqref{eq:preconditioning} suggested in~\cite{karras2022elucidating}. $\cshift$ is changed from $\by$ to $\mathbf{0}$.}
    \mybelowcaptionskip
    \label{fig:experiment_a}
\end{figure}

Figure~\ref{fig:experiment_a} shows the results of the preconditioning experiment.
The speech enhancement performance is reported in terms of ${\dpesq}$, ${\destoi}$ and ${\dsnr}$ as a function of the number of sampling steps ${n_{\mathrm{steps}}}$.
The dashed line corresponds to the SGMSE+M baseline from~\cite{lemercier2023analysing} and uses the preconditioning shown in Eq.~\eqref{eq:sgmse_precond}.
It can be seen that setting the ${c_{\mathrm{noise}}}$ parameter from ${\log t}$ to ${\frac{1}{4} \log \sigmahat(t)}$ as suggested in EDM has a detrimental effect on performance.
Similarly, setting ${c_{\mathrm{shift}}}$ to zero has a slight detrimental effect on performance.
Subsequently setting ${c_{\mathrm{in}}}$, ${c_{\mathrm{skip}}}$ and ${c_{\mathrm{out}}}$ to the values suggested in EDM also reduces performance.
However, once ${w(t)}$ is set to the value suggested in EDM such that the full EDM preconditioning is used, the performance is recovered and is substantially higher than the baseline for all metrics and all ${n_{\mathrm{steps}}}$ on the MultiCorpus dataset.
For VBDMD, the performance is superior in terms of ${\dsnr}$ for all ${n_{\mathrm{steps}}}$, but remains slightly inferior to the baseline in terms of ${\dpesq}$ and ${\destoi}$ at large ${n_{\mathrm{steps}}}$.

\subsection{SDE}

\begin{figure}
    \centering
    \myabovesinglefigureskip
    \includegraphics{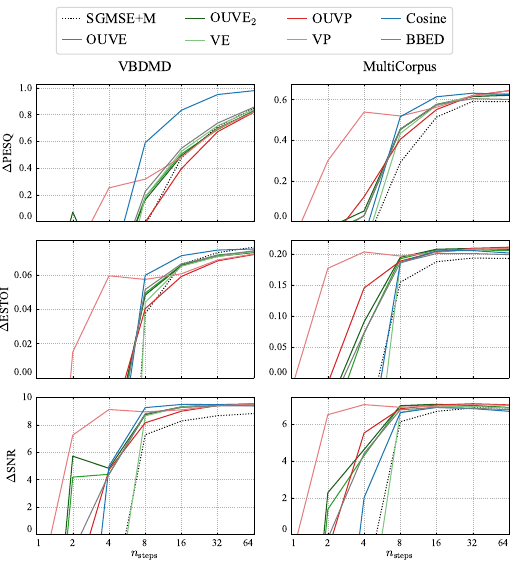}
    \myabovecaptionskip
    \caption{Speech enhancement performance as a function of the number of sampling steps ${n_{\mathrm{steps}}}$ for different \cglspl{sde}.}
    \mybelowcaptionskip
    \label{fig:experiment_b}
\end{figure}

Figure~\ref{fig:experiment_b} shows the results of the \cgls{sde} experiment.
It can be seen that the OUVE and OUVE\textsubscript{2} \cglspl{sde} show very similar performance.
This means that while we argue the OUVE\textsubscript{2} formulation is more practical, it does not lead to a performance improvement.
More interestingly, the \cgls{ve} \cgls{sde} shows similar performance to OUVE and OUVE\textsubscript{2} for ${n_{\mathrm{steps}} \inlinegeq 8}$.
Since the \cgls{ve} \cgls{sde} presents no drift of the diffusion process from the clean speech towards the noisy speech, this result suggests that the performance of diffusion-based speech enhancement cannot be attributed to such a drift as suggested in~\cite{welker2022speech,richter2023speech}.
Moreover, since the \cgls{ve} \cgls{sde} imposes a substantially larger prior mismatch compared to OUVE and OUVE\textsubscript{2}, this result also suggests that a prior mismatch is not necessarily responsible for a performance drop as suggested in~\cite{lay2023reducing}.
Similar conclusions can be drawn when comparing the \cgls{vp} and OUVP \cglspl{sde}: they show similar performance at large ${n_{\mathrm{steps}}}$, and the \cgls{vp} \cgls{sde} shows substantially higher performance at small ${n_{\mathrm{steps}}}$.
This means that in this particular example, the drift of the diffusion process from the clean speech towards the noisy speech actually has a detrimental effect on performance, despite the reduced prior mismatch.
The BBED \cgls{sde} shows similar performance to the OUVE and OUVE\textsubscript{2} \cglspl{sde} despite the minimized prior mismatch.
Finally, the cosine \cgls{sde} shows significantly higher ${\dpesq}$ scores on VBDMD.

\subsection{Stochasticity injected in reverse process}

\begin{figure*}[t]
    \centering
    \myabovedoublefigureskip
    \subfloat[\cgls{pc} sampler\label{fig:experiment_c_0}]{\includegraphics{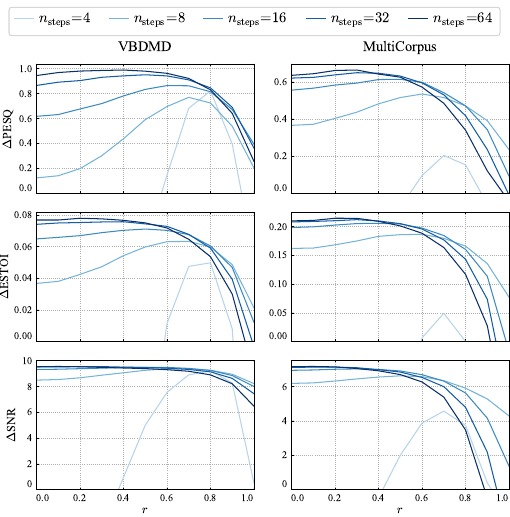}}
    \hfil
    \subfloat[Heun-based sampler\label{fig:experiment_c_1}]{\includegraphics{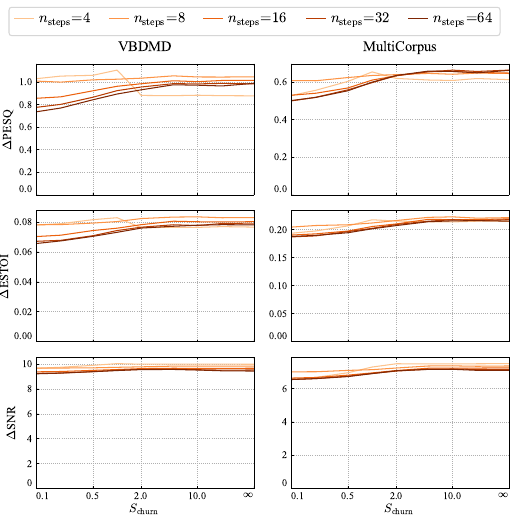}}
    \caption{Speech enhancement performance as a function of the amount of stochasticity injected in the reverse process for different numbers of sampling steps ${n_{\mathrm{steps}}}$.
    For the \cgls{pc} sampler (a), the stochasticity is controlled by the step size of the annealed Langevin dynamics correction ${r}$.
    For the Heun-based sampler (b), this is controlled by the parameter ${S_{\mathrm{churn}}}$.}
    \mybelowcaptionskip
    \label{fig:experiment_c_x}
\end{figure*}

Figure~\ref{fig:experiment_c_x} shows the results of the stochasticity experiment.
Figure~\ref{fig:experiment_c_0} shows the results for the \cgls{pc} sampler as a function of the step size of the annealed Langevin dynamics correction ${r}$, while Fig.~\ref{fig:experiment_c_1} shows the results for the Heun-based sampler as a function of the parameter ${S_{\mathrm{churn}}}$.
The experiment is repeated for different numbers of sampling steps ${n_{\mathrm{steps}}}$.
It can be seen that for the \cgls{pc} sampler, the step size ${r}$ has a large influence on performance.
When using a small ${n_{\mathrm{steps}}}$, a sufficient and precise amount of stochasticity is required, as the curves present sharp maxima across all metrics.
However, when increasing the number of sampling steps, the need for stochasticity is reduced, as the curves become flatter at low ${r}$ values.
Large ${r}$ values have a detrimental effect on performance in both cases.
For the Heun-based sampler, the parameter ${S_{\mathrm{churn}}}$ has a very small influence on performance.
Nonetheless, a slight rising trend can be observed, which suggests that injecting as much stochasticity as possible in the reverse process is beneficial for the Heun-based sampler.
However, as explained in Sect.~\ref{sec:sampler_exp}, the Heun-based sampler algorithm clamps the amount of noise injected in the reverse process, and we did not try removing this clamping.
In~\cite{karras2022elucidating}, it was suggested that while the need for stochasticity reduces as the model improves, models trained with diverse datasets continue to benefit from stochasticity.
Therefore, assuming these observations are valid for speech enhancement, our results suggest that the model can be further improved, or that the datasets are diverse enough.

\subsection{Comparison with baselines}

\begin{figure}
    \centering
    \myabovesinglefigureskip
    \includegraphics{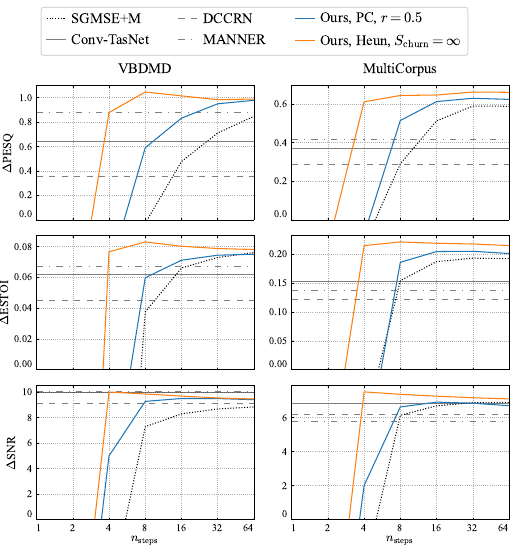}
    \myabovecaptionskip
    \caption{Speech enhancement performance as a function of ${n_{\mathrm{steps}}}$ for two different sampler configurations and the baselines.}
    \mybelowcaptionskip
    \label{fig:experiment_c}
\end{figure}

We set the preconditioning and the \cgls{sde} as in the previous experiment.
We chose between the \cgls{pc} sampler with ${r \inlineeq 0.5}$ as suggested in~\cite{richter2023speech} and the Heun-based sampler with ${S_{\mathrm{churn}} \inlineeq \infty}$.
Figure~\ref{fig:experiment_c} shows the performance of the two resulting systems and SGMSE+M as a function of the number of sampling steps ${n_{\mathrm{steps}}}$.
We also plot the performance of the discriminative baselines as horizontal lines.
When comparing our systems using both samplers, it can be seen that the Heun-based sampler shows superior performance compared to the \cgls{pc} sampler.
More specifically, the Heun-based sampler shows good performance on both datasets with only 4 sampling steps, while the \cgls{pc} sampler requires at least 16 sampling steps to achieve similar performance.
This means the Heun-based sampler allows to reduce the computational cost by a factor of four without losing performance.
When comparing with SGMSE+M, it can be seen that our system using the Heun-based sampler outperforms the baseline for all metrics and all ${n_{\mathrm{steps}}}$.
Finally, while the discriminative baselines are competitive in terms of ${\dsnr}$ on both datasets, the diffusion-based systems substantially outperform them in terms of ${\dpesq}$ and ${\destoi}$ when using large ${n_{\mathrm{steps}}}$ (except SGMSE+M which is outperformed by MANNER in terms of ${\dpesq}$ on VBDMD).

\subsection{Applying the Heun-based sampler to other SDEs}

To verify the benefit of the Heun-based sampler, we investigate if applying the Heun-based sampler to \cglspl{sde} different from the cosine \cgls{sde} also leads to a performance improvement.
We choose the OUVE \cgls{sde}, since it is the one used by the SGMSE+M baseline, and the VP \cgls{sde}, since it presents no drift of the diffusion process and it showed substantially higher performance at small ${n_{\mathrm{steps}}}$ in Fig.~\ref{fig:experiment_b}.
The results are shown in Tab.~\ref{tab:label} for ${n_{\mathrm{steps}} \inlineeq 4}$ and ${n_{\mathrm{steps}} \inlineeq 64}$.
When using ${n_{\mathrm{steps}} \inlineeq 4}$, SGMSE+M degrades the speech signal, as indicated by the negative scores.
Applying the Heun-based sampler to SGMSE+M (``+Heun'') substantially increases the scores, but the objective metric improvements are still very small.
It is only when the Heun-based sampler is applied in conjunction with the EDM preconditioning (``pre+Heun'') that the system substantially enhances the speech signal across all metrics.
Similarly, the Heun-based sampler substantially improves the performance when using the VP \cgls{sde} (``pre+VP'' vs.\ ``pre+VP+Heun'').
When using ${n_{\mathrm{steps}} \inlineeq 64}$, the Heun-based sampler is still beneficial for all configurations, although there is less variation in performance across the different configurations.
The system combining the EDM preconditioning, the cosine \cgls{sde} and the Heun-based sampler (``pre+cos+Heun'') shows the best performance in terms of ${\dpesq}$ and ${\destoi}$ on VBDMD, while the system combining the EDM preconditioning, the VP \cgls{sde} and the Heun-based sampler (``pre+VP+Heun'') shows the best performance in terms of ${\dpesq}$ on the MultiCorpus dataset.

\begin{table}
\scriptsize
\setlength{\tabcolsep}{2.5pt}
\centering
\caption{Speech enhancement performance for the baselines and for the diffusion-based approach using different preconditioning, \cgls{sde} and sampler configurations}
\label{tab:label}
\begin{tabular}{lcccccc}
\toprule
 & \multicolumn{3}{c}{VBDMD} & \multicolumn{3}{c}{MultiCorpus} \\
\cmidrule(lr){2-4} \cmidrule(lr){5-7}
 & \multicolumn{1}{c}{$\Delta$PESQ} & \multicolumn{1}{c}{$\Delta$ESTOI} & \multicolumn{1}{c}{$\Delta$SNR} & \multicolumn{1}{c}{$\Delta$PESQ} & \multicolumn{1}{c}{$\Delta$ESTOI} & \multicolumn{1}{c}{$\Delta$SNR} \\
\midrule
Conv-TasNet & \phantom{-}0.64{\tiny\:±\:0.43} & \phantom{-}0.06{\tiny\:±\:0.07} & \phantom{-}9.97{\tiny\:±\:4.30} & \phantom{-}0.37{\tiny\:±\:0.22} & \phantom{-}0.15{\tiny\:±\:0.08} & \phantom{-}6.84{\tiny\:±\:2.97} \\
DCCRN & \phantom{-}0.36{\tiny\:±\:0.37} & \phantom{-}0.05{\tiny\:±\:0.05} & \phantom{-}9.11{\tiny\:±\:4.00} & \phantom{-}0.29{\tiny\:±\:0.20} & \phantom{-}0.12{\tiny\:±\:0.05} & \phantom{-}6.15{\tiny\:±\:2.67} \\
MANNER & \phantom{-}0.88{\tiny\:±\:0.45} & \phantom{-}0.07{\tiny\:±\:0.07} & \phantom{-}\bfseries 9.99{\tiny\:±\:4.23} & \phantom{-}0.42{\tiny\:±\:0.25} & \phantom{-}0.14{\tiny\:±\:0.06} & \phantom{-}5.76{\tiny\:±\:2.57} \\
\midrule$n_{\mathrm{steps}}=4$ & & & & & & \\\cmidrule{1-1}
SGMSE+M & -0.90{\tiny\:±\:0.74} & -0.27{\tiny\:±\:0.11} & -6.08{\tiny\:±\:5.34} & -0.13{\tiny\:±\:0.14} & -0.10{\tiny\:±\:0.09} & -3.06{\tiny\:±\:3.45} \\
+Heun & \phantom{-}0.02{\tiny\:±\:0.73} & \phantom{-}0.00{\tiny\:±\:0.10} & \phantom{-}3.05{\tiny\:±\:4.49} & \phantom{-}0.15{\tiny\:±\:0.19} & \phantom{-}0.07{\tiny\:±\:0.09} & \phantom{-}4.49{\tiny\:±\:2.72} \\
+pre & -0.56{\tiny\:±\:0.70} & -0.06{\tiny\:±\:0.08} & \phantom{-}4.41{\tiny\:±\:4.91} & \phantom{-}0.03{\tiny\:±\:0.12} & \phantom{-}0.07{\tiny\:±\:0.08} & \phantom{-}4.33{\tiny\:±\:3.00} \\
+pre+Heun & \phantom{-}0.71{\tiny\:±\:0.53} & \phantom{-}0.08{\tiny\:±\:0.07} & \phantom{-}9.90{\tiny\:±\:4.44} & \phantom{-}0.67{\tiny\:±\:0.32} & \phantom{-}0.22{\tiny\:±\:0.08} & \phantom{-}\bfseries 7.68{\tiny\:±\:2.68} \\
+pre+cos & -0.42{\tiny\:±\:0.66} & -0.09{\tiny\:±\:0.07} & \phantom{-}4.99{\tiny\:±\:4.70} & -0.08{\tiny\:±\:0.12} & -0.16{\tiny\:±\:0.10} & \phantom{-}2.01{\tiny\:±\:3.22} \\
+pre+VP & \phantom{-}0.25{\tiny\:±\:0.58} & \phantom{-}0.06{\tiny\:±\:0.07} & \phantom{-}9.11{\tiny\:±\:4.59} & \phantom{-}0.54{\tiny\:±\:0.27} & \phantom{-}0.20{\tiny\:±\:0.08} & \phantom{-}7.06{\tiny\:±\:2.82} \\
+pre+cos+Heun & \phantom{-}0.88{\tiny\:±\:0.43} & \phantom{-}0.08{\tiny\:±\:0.07} & \phantom{-}9.98{\tiny\:±\:4.50} & \phantom{-}0.61{\tiny\:±\:0.32} & \phantom{-}0.21{\tiny\:±\:0.09} & \phantom{-}7.51{\tiny\:±\:2.84} \\
+pre+VP+Heun & \phantom{-}0.75{\tiny\:±\:0.50} & \phantom{-}0.08{\tiny\:±\:0.07} & \phantom{-}9.85{\tiny\:±\:4.62} & \phantom{-}0.63{\tiny\:±\:0.35} & \phantom{-}\bfseries 0.22{\tiny\:±\:0.09} & \phantom{-}7.63{\tiny\:±\:2.70} \\
\midrule$n_{\mathrm{steps}}=64$ & & & & & & \\\cmidrule{1-1}
SGMSE+M & \phantom{-}0.85{\tiny\:±\:0.51} & \phantom{-}0.08{\tiny\:±\:0.07} & \phantom{-}8.82{\tiny\:±\:4.62} & \phantom{-}0.59{\tiny\:±\:0.38} & \phantom{-}0.19{\tiny\:±\:0.10} & \phantom{-}6.89{\tiny\:±\:2.92} \\
+Heun & \phantom{-}0.87{\tiny\:±\:0.49} & \phantom{-}0.08{\tiny\:±\:0.07} & \phantom{-}7.89{\tiny\:±\:4.40} & \phantom{-}0.57{\tiny\:±\:0.37} & \phantom{-}0.19{\tiny\:±\:0.10} & \phantom{-}6.48{\tiny\:±\:2.84} \\
+pre & \phantom{-}0.81{\tiny\:±\:0.48} & \phantom{-}0.07{\tiny\:±\:0.07} & \phantom{-}9.39{\tiny\:±\:4.48} & \phantom{-}0.62{\tiny\:±\:0.29} & \phantom{-}0.20{\tiny\:±\:0.09} & \phantom{-}6.87{\tiny\:±\:2.78} \\
+pre+Heun & \phantom{-}0.88{\tiny\:±\:0.49} & \phantom{-}0.08{\tiny\:±\:0.07} & \phantom{-}9.52{\tiny\:±\:4.48} & \phantom{-}0.66{\tiny\:±\:0.32} & \phantom{-}0.22{\tiny\:±\:0.09} & \phantom{-}7.09{\tiny\:±\:2.75} \\
+pre+cos & \phantom{-}0.98{\tiny\:±\:0.44} & \phantom{-}0.08{\tiny\:±\:0.07} & \phantom{-}9.38{\tiny\:±\:4.52} & \phantom{-}0.62{\tiny\:±\:0.32} & \phantom{-}0.20{\tiny\:±\:0.09} & \phantom{-}6.69{\tiny\:±\:2.82} \\
+pre+VP & \phantom{-}0.82{\tiny\:±\:0.46} & \phantom{-}0.07{\tiny\:±\:0.07} & \phantom{-}9.52{\tiny\:±\:4.55} & \phantom{-}0.64{\tiny\:±\:0.31} & \phantom{-}0.21{\tiny\:±\:0.09} & \phantom{-}7.02{\tiny\:±\:2.84} \\
+pre+cos+Heun & \phantom{-}\bfseries 0.99{\tiny\:±\:0.45} & \phantom{-}\bfseries 0.08{\tiny\:±\:0.07} & \phantom{-}9.44{\tiny\:±\:4.48} & \phantom{-}0.66{\tiny\:±\:0.35} & \phantom{-}0.22{\tiny\:±\:0.09} & \phantom{-}7.09{\tiny\:±\:2.80} \\
+pre+VP+Heun & \phantom{-}0.93{\tiny\:±\:0.46} & \phantom{-}0.08{\tiny\:±\:0.07} & \phantom{-}9.54{\tiny\:±\:4.54} & \phantom{-}\bfseries 0.68{\tiny\:±\:0.33} & \phantom{-}0.22{\tiny\:±\:0.09} & \phantom{-}7.10{\tiny\:±\:2.78} \\
\bottomrule
\multicolumn{7}{l}{Values indicate mean and standard deviation across mixtures in the test dataset.} \\
\multicolumn{7}{l}{Values in bold indicate the best mean performance for each column.}
\end{tabular}
\end{table}

\section{Conclusion}
\label{sec:conclusion}

In the present work, we extended the EDM framework proposed in~\cite{karras2022elucidating} to include a recent diffusion-based speech enhancement model~\cite{welker2022speech,richter2023speech}.
Specifically, we considered a more general \cgls{sde} with a non-zero long-term mean, we added a new parameter ${c_{\mathrm{shift}}}$ to the neural network preconditioning, and we showed that the score function can be evaluated on the unshifted and unscaled process instead of the unscaled process.
This allowed to shed light on the different design aspects of diffusion models in the case of speech enhancement, and to apply recent developments from image generation literature that were not investigated for speech enhancement before.

By experimenting with \cglspl{sde} that were not used for speech enhancement before, we were able to show that similar performance can be obtained with \cglspl{sde} that present a drift from the clean speech towards the noisy speech compared to \cglspl{sde} that do not present such a drift.
This suggests that the model does not require such a drift to reconstruct speech corrupted by environmental noises that differ from the stationary Gaussian noise of the diffusion process, contrary to previous beliefs~\cite{richter2023speech}.
This similar performance was observed despite the increased mismatch between the final distribution of the forward diffusion process and the prior distribution used to initialize the reverse process, even though this was argued to be responsible for a drop in performance~\cite{lay2023reducing}.
Omitting the drift term makes the speech enhancement task fully generative, since this models a progressive transformation from Gaussian noise to clean speech, rather than from noisy speech to clean speech.
Moreover, we hypothesize that this improves training efficiency, since the network does not need to learn to undo the scaling ${s(t)}$ before denoising.

The usage of the Heun-based sampler in conjunction with the neural network preconditioning from~\cite{karras2022elucidating} led to a substantial improvement of speech enhancement performance when using few sampling steps.
More specifically, the Heun-based sampler required as few as 4 sampling steps to perform similarly to the \cgls{pc} sampler with 16 sampling steps, thus allowing for a fourfold decrease in computational cost.
The Heun-based sampler was also shown to be beneficial for different \cglspl{sde}.

Despite the encouraging results, diffusion-based speech enhancement is still computationally more costly compared to state-of-the-art discriminative systems due to the multiple neural network evaluations in the reverse process.
This raises the question of whether diffusion-based speech enhancement is worth the additional computational cost.
While further reducing the computational cost is an obvious direction for future work, other options include systematically investigating the performance in mismatched conditions, or exploring the benefit of an unsupervised training procedure for speech enhancement.
Indeed, the present work only investigated the performance in matched conditions, and while our previous study~\cite{gonzalez2024diffusion} reported results in mismatched conditions using multiple speech, noise and \cgls{brir} databases, this was not done in conjunction with the different design aspects of the diffusion model.
It is thus unclear if some specific~\cglspl{sde} or the stochasticity in the reverse process are beneficial to generalizing to unseen acoustic conditions.
Moreover, while the proposed diffusion-based approach is generative in that it learns a conditional probability distribution over clean speech, the models were trained in a supervised setting using pairs of clean and noisy speech signals.
It would be interesting to investigate if an unsupervised training procedure improves the robustness to arbitrary additive and convolutional distortions.

\appendices
\allowdisplaybreaks

\section{Cosine schedule parameters}
\label{app:cosine_schedule}

Under the \cgls{vp} assumption, we have ${\squared{s}{t} \inlineplus \squared{\sigma}{t} \inlineeq 1}$.
Therefore, using Eqs.~\eqref{eq:diffusion_snr} and \eqref{eq:cosine_schedule},
\begin{align}
    & \textstyle \squared{s}{t} + e^{-\lambda(t)} \squared{s}{t} = 1 \\
    \textstyle \implies & \textstyle \squared{s}{t} = \frac{1}{1 + e^{-\lambda(t)}} = \frac{1}{1 + e^{-2 \nu}\tansq \frac{\pi t}{2}}. \label{eq:s2_cosine}
\end{align}
For ${\sigma(t)}$, we can reuse ${\squared{s}{t} \inlineplus \squared{\sigma}{t} \inlineeq 1}$ such that
\begin{align}
    \textstyle \squared{\sigma}{t} & \textstyle = 1 - \squared{s}{t} = 1 - \frac{1}{1 + e^{-2 \nu}\tansq \frac{\pi t}{2}} \\
    & \textstyle = \frac{e^{-2 \nu}\tansq \frac{\pi t}{2}}{1 + e^{-2 \nu}\tansq \frac{\pi t}{2}} = \frac{1}{1 + e^{2 \nu}\cotsq \frac{\pi t}{2}}. \label{eq:sigma2_cosine}
\end{align}
The drift and diffusion coefficients ${f(t)}$ and ${g(t)}$ can be derived using Eqs.~\eqref{eq:revert_marginals}, \eqref{eq:s2_cosine} and \eqref{eq:sigma2_cosine}.
Specifically for ${f(t)}$,
\begin{align}
    \textstyle f(t) & \textstyle = \frac{\diff}{\diff t} \big[ \log s(t) \big] = \frac{\diff}{\diff t} \left[ \log \sqrt{\frac{1}{1 + e^{-2 \nu} \tansq \frac{\pi t}{2}}} \right] \\
    & \textstyle = -\frac{1}{2} \cdot \frac{\diff}{\diff t} \left[ \log \left( 1 + e^{-2 \nu}\tansq \frac{\pi t}{2} \right) \right] \\
    & \textstyle = -\frac{1}{2} \cdot \frac{e^{-2 \nu}}{1 + e^{-2 \nu}\tansq \frac{\pi t}{2}} \cdot \frac{\diff}{\diff t} \left[ \tansq \frac{\pi t}{2} \right] \\
    & \textstyle = -\frac{1}{2} \cdot \frac{e^{-2 \nu}}{1 + e^{-2 \nu}\tansq \frac{\pi t}{2}} \cdot \frac{\pi \tan \! \frac{\pi t}{2}}{\cossq \frac{\pi t}{2}} \\
    & \textstyle = -\frac{1}{2} \cdot \frac{e^{-2 \nu}\tansq \frac{\pi t}{2}}{1 + e^{-2 \nu}\tansq \frac{\pi t}{2}} \cdot \frac{\pi}{\cos \! \frac{\pi t}{2}\sin \! \frac{\pi t}{2}} \\
    & \textstyle = -\frac{1}{1 + e^{2 \nu}\cotsq \frac{\pi t}{2}} \cdot \frac{\pi}{\sin (\pi t)} = -\frac{\pi \csc(\pi t)}{1 + e^{2 \nu}\cotsq \frac{\pi t}{2}}.
\end{align}
For ${g(t)}$, we can use the fact that under the \cgls{vp} assumption, we have ${\squared{g}{t} \inlineeq -2f(t)}$ from Eq.~\eqref{eq:vp_beta_coeff}.
Therefore,
\begin{equation}
    \textstyle g(t) = \sqrt{\frac{2 \pi \csc(\pi t)}{1 + e^{2 \nu}\cotsq \frac{\pi t}{2}}}.
\end{equation}

\section{Proof of Eqs.~\texorpdfstring{\eqref{eq:denoiser_loss_2}}{(43)} and \texorpdfstring{\eqref{eq:sgmse_precond}}{(44)}}
\label{app:sgmse_precond}

We denote the unscaled noise as ${\unscalednoise \inlineeq \sigmahat(t) \bz}$ and the scaled noise as ${ \scalednoise \inlineeq \sigma(t) \bz \inlineeq s(t) \sigmahat(t) \bz \inlineeq s(t) \unscalednoise}$ where ${\bz \inlinesim \gaussian_\C(0, \verythinnegativespace \mathbf{I})}$.
Similarly to~\cite{karras2022elucidating}, we start from the term inside the expectation operator in Eq.~\eqref{eq:sgmse_loss} and use the expression of the score model as a function of the raw neural network layers in Eq.~\eqref{eq:sgmse_score_model},
\begin{flalign}
    &\scalebox{.97}{$
        \left\| \sigma(t) \scoremodel(\bx_t, \by, \log t) \inlineplus \bz \right\|_2^2
        \inlineeq \left\| -\frac{\sigma(t)}{t} \model(\bx_t, \by, \log t) \inlineplus \bz \right\|_2^2
    $}\!\!\!\!\!\!\!&\\
    &\!=\scalebox{1}{$
        \left\| -\frac{\sigma(t)}{t} \model \big( s(t)(\bx_0 - \by) + \by + \scalednoise, \by, \log t \big) + \bz \right\|_2^2
    $}\!\!\!\!\!\!\!\\
    &\!=\scalebox{1}{$
        \left\| -\frac{\sigma(t)}{t} \model \big( s(t)\bxtilde_0 + \by + s(t)\unscalednoise, \by, \log t \big) + \bz \right\|_2^2
    $}\!\!\!\!\!\!\!\\
    &\!=\scalebox{.96}{$
        \frac{1}{\squared{\sigmahat}{t}} \! \left\| -\frac{s(t) \squared{\sigmahat}{t}}{t} \model \big( s(t)(\bxtilde_0 \inlineplus \unscalednoise) \inlineplus \by,
        \by, \log t \big) + \unscalednoise \right\|_2^2
    $}\!\!\!\!\!\!\!\\
    &\!=\scalebox{.8}{$
        \frac{1}{\squared{\sigmahat}{t}} \! \left\| -\frac{s(t)\squared{\sigmahat}{t}}{t} \model \big( s(t)(\bxtilde_0 \inlineplus \unscalednoise) \inlineplus \by,
        \by, \log t \big) \inlineplus (\bxtilde_0 \inlineplus \unscalednoise) \inlineminus \bxtilde_0 \right\|_2^2
    $}\!\!\!\!\!\!\!\\
    &\!=\scalebox{1}{$
        \smallunderbrace{\textstyle \frac{1}{\squared{\sigmahat}{t}}}_{w(t)} \left\| \denoiser(\bxtilde_0 + \unscalednoise, \by, t) - \bxtilde_0 \right\|_2^2,
    $}\!\!\!\!\!\!\!
\end{flalign}
where ${w(t)}$ and the preconditioning become apparent,
\begin{equation}
    \resizebox{.89\hsize}{!}{$
    \displaystyle
    \denoiser(\bx, \by, t) = \\
    \!\!\! \smallunderbrace{1}_{\cskip} \!\!
    \bx \smallunderbrace{\textstyle-\frac{s(t)\squared{\sigmahat}{t}}{t}}_{\cout}
    \model \big( \smallunderbrace{s(t)}_{\cin} \bx +
    \!\!\! \smallunderbrace{\by}_{\cshift} \!\!
    , \by, \smallunderbrace{\log t}_{\cnoise} \big).
    $}
\end{equation}

\section{Proof of Eq.~\texorpdfstring{\eqref{eq:sampling_eq}}{(45)}}
\label{app:sampling_eq}
Starting from the definition of the marginal distribution and following similar steps as in Eqs.~(15-19) in~\cite{karras2022elucidating},
\begin{align}
    &p_t(\bx_t | \by) = \int_{\R^d} \perturbkernelsgmse \pdata(\bx_0) \diff \bx_0 \\
    &\!\!= \int_{\R^d} \gaussian \big( \bx_t ; s(t)(\bx_0 - \by) + \by, \squared{\sigma}{t} \mathbf{I} \big) \pdata(\bx_0) \diff \bx_0 \\
    &\!\!= \int_{\R^d} \gaussian \big( \bx_t -\by ; s(t)(\bx_0 - \by), \squared{\sigma}{t} \mathbf{I} \big) \pdata(\bx_0) \diff \bx_0 \\
    &\!\!= \int_{\R^d} \gaussian \big( \bxtilde_t ; s(t) \bxtilde_0, \squared{\sigma}{t} \mathbf{I} \big) \pdata(\bxtilde_0 + y) \diff \bxtilde_0 \\
    &\!\!= s(t)^{-d} \int_{\R^d} \gaussian \left(\frac{\bxtilde_t}{s(t)} ; \bxtilde_0, \squared{\sigmahat}{t} \mathbf{I}\right) \pdata(\bxtilde_0 + y) \diff \bxtilde_0 \\
    &\!\!= s(t)^{-d} \Big[ \ptildedata \ast \gaussian \big( \mathbf{0}, \squared{\sigmahat}{t} \mathbf{I} \big) \Big] \left( \frac{\bxtilde_t}{s(t)} \right),
\end{align}
where ${\ptildedata : \bx \mapsto \pdata(\bx \inlineplus \by)}$ is a shifted version of ${\pdata}$ and ${\ast}$ is the convolution operator between \cglspl{pdf}.

\bibliographystyle{IEEEtran}
\bibliography{IEEEabrv, abbrv, refs}

\end{document}